\pdfoutput=1
\documentclass[atlasdraft=false, texlive=2023, UKenglish, cernpreprint, texmf, orcidlogo]{atlasdoc}

\usepackage{atlaspackage}
\usepackage{atlasbiblatex}
 
\usepackage{atlasphysics}

\graphicspath{{logos/}{figures/}}
 
\usepackage{ANA-HDBS-2018-50-PAPER-defs}

\AtlasTitle{Search for dark matter produced in association with a
Higgs boson decaying to tau leptons at $\sqrt{s}=13$~TeV with
the ATLAS detector}

\AtlasAbstract{
A search for dark matter produced in
association with a Higgs boson in
final states with two hadronically decaying $\tau$-leptons and missing
transverse momentum is
presented. The analysis uses $139~\mathrm{fb^{-1}}$ of proton--proton
collision data at $\sqrt{s}=13$~TeV collected by the ATLAS experiment
at the Large Hadron Collider between 2015 and 2018.
No evidence of physics beyond the Standard Model is found. The
results are interpreted in terms of a 2HDM+\textit{a} model featuring two scalar Higgs doublets and a pseudoscalar singlet field. Exclusion limits on the parameters of the model in selected benchmark scenarios are derived at 95\% confidence level.
Model-independent limits are also set on the visible cross-section for
processes beyond the Standard Model producing missing transverse
momentum in association with a Higgs boson decaying into $\tau$-leptons.
}

\AtlasRefCode{HDBS-2018-50}
\PreprintIdNumber{CERN-EP-2023-072}

\AtlasJournal{JHEP}
\AtlasJournalRef{JHEP 09 (2023) 189}
\AtlasDOI{10.1007/JHEP09(2023)189}

\hypersetup{pdftitle={ATLAS document},pdfauthor={The ATLAS Collaboration}}
 
\begin{document}
 
\maketitle
 
\tableofcontents

\FloatBarrier
\section{Introduction}

The Standard Model (SM) of particle physics is a successful and experimentally validated theory describing elementary particles and their interactions, culminating with the discovery in 2012 \cite{HIGG-2012-27,CMS-HIG-12-028} of a particle consistent with the Higgs boson ($h$) predicted in the SM. However, in its current form, the SM does not contain any dark matter (DM) particle candidate, whose existence is supported by astrophysical observations \cite{Clowe:2003tk,Bertone:2004pz,ParticleDataGroup:2020ssz}. This provides a strong motivation for searches for physics beyond the Standard Model (BSM physics), in particular DM candidates \cite{Kahlhoefer:2017dnp}.
 
During the \RunTwo data-taking period, the Large Hadron Collider (LHC) experiments collected a substantial dataset, which can be used to search for signs of DM. There is no evidence of non-gravitational interactions between the DM and SM particles; this makes direct observation with general-purpose detectors such as ATLAS \cite{PERF-2007-01} or CMS \cite{CMS-CMS-00-001} unlikely. The majority of the candidate DM particles $\chi$ would not interact with the material of the detector, and would escape undetected. One alternative strategy is to focus on the case of DM particles produced in association with SM particles, focusing on topologies where a single SM particle X is produced \cite{Nachtmann:1984xu}.
Such events do not rely on any model-specific assumptions and are characterised by a transverse-momentum imbalance ($\met$) due to the DM particle escaping the detector, resulting in \enquote{X+$\met$} signatures, also referred to as \enquote{mono-X}.
 
The visible SM particle X can be a jet, a photon, a $W$ or $Z$ boson, a
top quark, or a Higgs boson \cite{Abercrombie:2015wmb}.  
Typically, the couplings of such particles to light quarks and gluons of the SM are much stronger than their coupling to DM particles. The one obvious exception is a Higgs boson $h$ whose couplings to light quarks and gluons are suppressed. This means that the mono-Higgs topologies are only significantly sensitive to scenarios in which the Higgs boson couples directly to some BSM particle participating in DM production. In such cases the direct SM--BSM interactions are probed, making the mono-Higgs searches a window into the structure of BSM physics responsible for DM production. In many theories where the Higgs couplings to BSM particles are enhanced \cite{GAMBIT:2018eea}, such scenarios are possible, making mono-Higgs topologies an important tool in LHC DM searches.
 
The ATLAS and CMS collaborations have performed $h+\met$ searches where
the Higgs boson decays into a pair of
$b$-quarks \cite{EXOT-2018-46,CMS-EXO-16-050},
$\tau$-leptons\footnote{Direct comparison with the presented search is not possible due to the difference between the models used for reinterpretation.}
\cite{CMS-EXO-16-055}, or photons \cite{HIGG-2019-02,CMS-EXO-16-012}, or into $ZZ$ or $W^+W^-$ \cite{CMS-EXO-18-011}.
This analysis presents a search for DM using the $h + \met$ topology with the Higgs boson decaying into two hadronically decaying $\tau$-leptons, using the $139$~fb$^{-1}$ of proton--proton collision data collected by the ATLAS detector in 2015--2018 at $\sqrt{s} = 13$~\TeV. The presence of tau neutrinos can, in certain topologies, enhance the \met\ signature of the signal and increase the sensitivity to the more challenging models ~\cite{SUSY-2018-31}.

\begin{figure}[t]
\centering
\subfloat[]{\includegraphics[width= 0.3\textwidth]{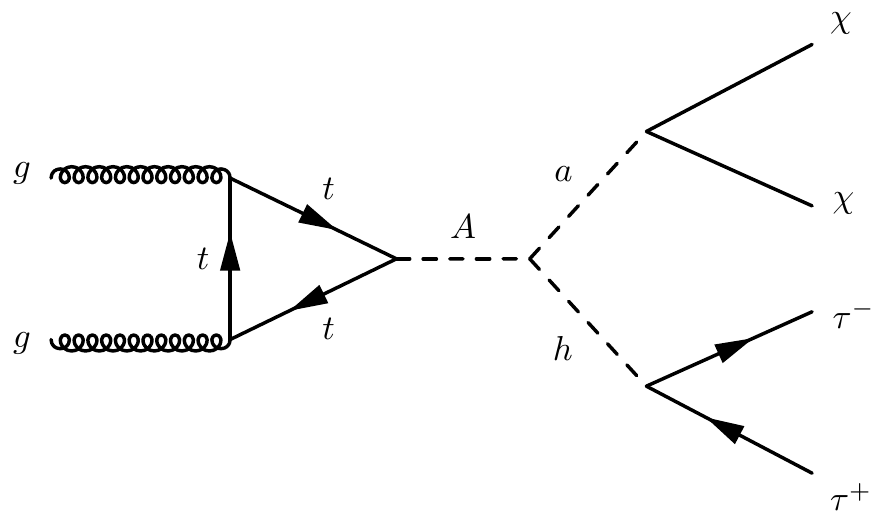}\label{fig:feynman_ggF}}
\hfil
\subfloat[]{\includegraphics[width= 0.3\textwidth]{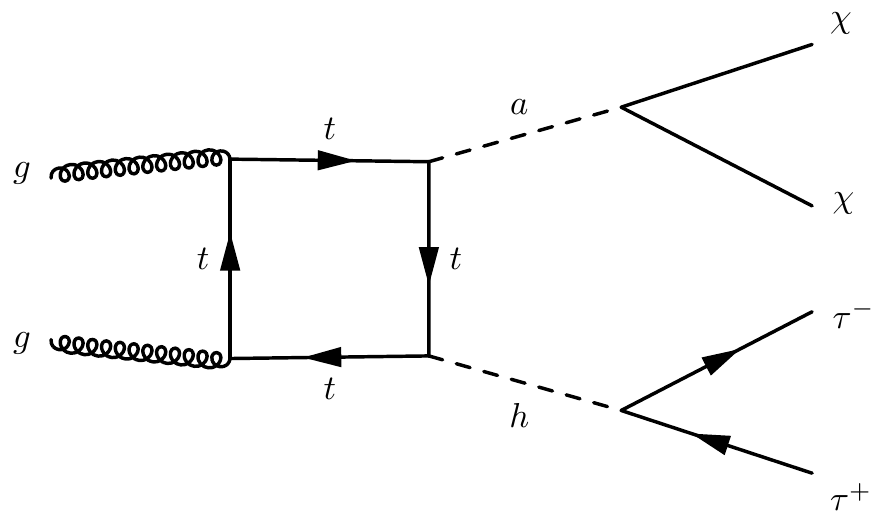}\label{fig:feynman_ggF2}}
\hfil
\subfloat[]{\includegraphics[width= 0.3\textwidth]{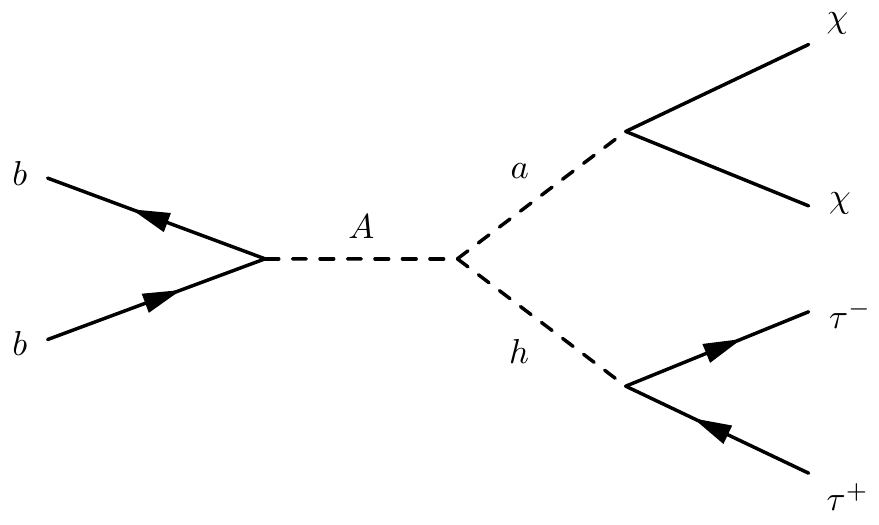}\label{fig:feynman_bb}}
 
\caption{Feynman diagrams describing the \TwoHDMa scenarios leading to the production of the mono-Higgs signatures: \protect\subref{fig:feynman_ggF} and \protect\subref{fig:feynman_ggF2} gluon--gluon fusion, \protect\subref{fig:feynman_bb} $b\bar{b}$ annihilation.}
\label{fig:feynm_diag}
\end{figure}

One of the most straightforward and natural extensions of the SM is
the two-Higgs-doublet model (2HDM)~\cite{PhysRevD.18.2574}, extending the Higgs sector with a second scalar doublet.
Such an extension of the Higgs sector is required in supersymmetric
models, but is also of more general interest~\cite{Branco:2011iw}. After electroweak symmetry breaking the model has five physical Higgs states: two neutral scalar states $h$ and $H$, a pseudoscalar neutral state $A$, and two charged scalar states $H^{\pm}$. The \TwoHDMa model \cite{Bauer:2017ota} extends the baseline 2HDM by adding a new pseudoscalar singlet $a$ with Yukawa-like coupling to the DM candidate $\chi$ and the SM fermions. The singlet acts as a mediator between the SM particles and the DM sector. This is the simplest gauge-invariant and renormalisable extension of pseudoscalar mediator DM models \cite{Abe:2644694}. The model leads to $h + \met$ final states through the gluon--gluon fusion and $b\bar{b}$ annihilation modes as depicted by the Feynman diagrams of Figure \ref{fig:feynm_diag}.

The present analysis considers the \TwoHDMa model as a benchmark and
uses it to optimise the search and interpret the results. The model is
described by 12 additional parameters not present in the SM.
Following the LHC DM Working Group's recommendations \cite{Abe:2644694}, two-dimensional parameter scans are performed
with the following benchmark parameter choices:
 
\begin{itemize}
\item Masses of the CP-odd Higgs boson $A$, CP-even Higgs boson $H$, and charged Higgs bosons $H^{\pm}$ are set to be equal to each other.
\item The mass of the fermionic DM particle $\chi$ is set to $10$~\GeV.
\item The DM Yukawa coupling $y_{\chi}$ to pseudoscalar $a$ is set to 1.
\item Mixing angles $\alpha$ and $\beta$ are constrained to satisfy $\cos(\beta - \alpha) = 0$, where $\alpha$ is the mixing angle between CP-even Higgs bosons $h$ and $H$, and $\tan \beta$ is the ratio of the vacuum expectation values (VEV) of the two Higgs doublets. In this so-called alignment limit the properties of the lightest Higgs boson $h$ are very similar to those of the SM Higgs boson.
\item Quartic couplings of the pseudoscalar potentials and Higgs potential $\lambda_\text{1P}$, $\lambda_\text{2P}$ and $\lambda_3$ are all set to 3.
\item The mass of the lightest Higgs boson $h$ (similar to the SM Higgs boson) and the value of EW VEV $v$ are fixed by observations to be $m_h \approx 125$~\GeV~\cite{HIGG-2014-14} and $v \approx 246$~\GeV\  \cite{Tanabashi:2636832}.
\item The mixing angle between the two pseudoscalars, $\Theta$, is required to satisfy  $\sin \Theta = 0.35$ or $\sin \Theta = 0.7$.
\end{itemize}
 
The analysis considers two two-dimensional scans in parameter values -- the first in $m_A$ and $m_a$, the second in $m_A$ and $\tan \beta$. The kinematic dependence of the signal model on $m_A$ in particular is non-trivial and the analysis is optimised for different ranges of $m_A$ (\enquote{low} and \enquote{high}).

SM processes can also produce signatures with $\met$ and two $\tau$-leptons consistent with the decay products of a Higgs boson. The production of a Higgs boson in association with a $Z$ boson decaying into neutrinos is of particular importance as, despite the relatively low cross-section, it can lead to final states indistinguishable from the signal. The production of a $Z$ boson, a pair of vector bosons, \ttbar, or a Higgs boson in association with a $W$ boson can also lead to final states with \met and two $\tau$-leptons. Due to the larger cross-sections, finite detector resolution and acceptance, and the ambiguity introduced by the presence of two neutrinos, these processes can mimic the potential signal events.

The paper is structured as follows.
Section~\ref{sec:detector} gives a short description of the ATLAS
detector. Section~\ref{sec:samples} provides information about the
dataset and Monte Carlo simulations used by the
analysis. Section~\ref{sec:reconstruction} describes the various physical
objects used in the analysis. In Section~\ref{sec:selection}, the
selections applied to the dataset are summarised. Sections~\ref{sec:background} and~\ref{sec:systematics} describe the modelling and systematic uncertainties, respectively, of signal and SM backgrounds. The results of the analysis and the interpretations are discussed in Section~\ref{sec:result}, with a short summary given in Section~\ref{sec:conclusion}.


\FloatBarrier

\section{ATLAS detector}
\label{sec:detector}

The ATLAS experiment~\cite{PERF-2007-01,ATLAS-TDR-2010-19,PIX-2018-001} is a multipurpose cylindrical particle detector at the LHC with nearly \(4\pi\) coverage in
solid angle.\footnote{ATLAS uses a right-handed coordinate system with its origin at the nominal interaction point (IP)
in the centre of the detector and the \(z\)-axis along the beam pipe.
The \(x\)-axis points from the IP to the centre of the LHC ring,
and the \(y\)-axis points upwards.
Cylindrical coordinates \((r,\phi)\) are used in the transverse plane,
\(\phi\) being the azimuthal angle around the \(z\)-axis.
The pseudorapidity is defined in terms of the polar angle \(\theta\) as \(\eta = -\ln \tan(\theta/2)\).
Angular distance is measured in units of \(\Delta R \equiv \sqrt{(\Delta\eta)^{2} + (\Delta\phi)^{2}}\).
}

At the centre of the ATLAS detector is the inner tracking detector (ID). It is surrounded by a thin superconducting solenoid
providing a \SI{2}{\tesla} axial magnetic field. The ID consists of silicon pixel, silicon microstrip, and transition radiation tracking detectors and covers the pseudorapidity range \(|\eta| < 2.5\).
The next layer of the ATLAS detector is made of lead/liquid-argon
(LAr) sampling calorimeters that measure the energy and position of
electromagnetic showers with high granularity. In the central
region \(|\eta| < 1.7\) the steel/scintillator-tile calorimeter is
used to provide hadronic energy measurements. In the endcap and
forward regions, LAr calorimeters with copper or tungsten absorbers
are used for electromagnetic and hadronic shower measurements up
to \(|\eta| = 4.9\). The outermost layer of the ATLAS detector is the muon
spectrometer. It consists of separate trigger and high-precision
tracking chambers, operating in a magnetic field generated by three large air-core toroidal superconducting magnets with eight coils each.

To select and record interesting events, a two-level trigger system is used. The level-1 trigger is implemented in custom hardware. It uses information from the calorimeters and the muon spectrometer
to reduce the event rate to below \SI{100}{\kHz}. This is followed by the high-level trigger (HLT), which is fully software-based. It is used to further reduce the event rate to  \SI{1}{\kHz} on average.
 
An extensive software suite~\cite{ATL-SOFT-PUB-2021-001} is used in data simulation, in the reconstruction and analysis of real and simulated data, in detector operations, and in the trigger and data acquisition systems of the experiment.


\section{Data and simulated event samples}
\label{sec:samples}


The dataset used by the analysis was collected with the ATLAS detector during proton--proton collisions at a centre-of-mass energy $\sqrt{s} = 13$~\TeV\ during \RunTwo of the LHC (2015--2018). The recorded data correspond to an integrated luminosity of $139~\mathrm{fb}^{-1}$. A set of data-quality requirements are applied to the dataset to ensure that the LHC beam conditions were stable and the ATLAS detector was fully functional \cite{DAPR-2018-01}.
All data events considered by this
analysis are required to pass a combined di-$\tauhadvis$ + \met\ trigger
~\cite{TRIG-2016-01,ATLAS-CONF-2017-061},
where $\tauhadvis$ is the visible part of a hadronically decaying $\tau$-lepton.

The analysis uses Monte Carlo (MC) simulation to model \TwoHDMa signal events and the majority of SM the backgrounds. The ATLAS simulation framework \cite{SOFT-2010-01} was used to produce all of the MC samples. A full detector-response simulation based on \GEANT~\cite{Agostinelli:2002hh} is used. This section describes the technical details of the MC samples, which are also summarised in Table~\ref{tab:samples_desc}.

\begin{table}[!h]
\tiny
\centering
\caption{Summary of the generators used to simulate the background processes considered in the analysis. The \enquote{dedicated} tune refers to the set of tuned parton-shower parameters developed by the \SHERPA authors. The symbol \enquote{$V$} stands for a vector boson.}
\label{tab:samples_desc}
\begin{tabular}{lp{0.18\linewidth}p{0.143\linewidth}p{0.123\linewidth}p{0.059\linewidth}p{0.104\linewidth}p{0.11\linewidth}}
\toprule
 
Physics process          			& Generator 	& Parton Shower 	&  Accuracy 		& Tune 			& PDF (generator) & PDF (PS)\\
\midrule
$Z(\rightarrow \ell\ell )$+jets    & \SHERPA[2.2.1]~\cite{Bothmann:2019yzt}                        & \SHERPA[2.2.1]~\cite{Schumann:2007mg}         & NNLO~\cite{Anastasiou:2003ds}                 & Dedicated       & \NNPDF[3.0nnlo]~\cite{Ball:2014uwa} & \NNPDF[3.0nnlo]~\cite{Ball:2014uwa} \\
$W(\rightarrow \ell\nu )$+jets    & \SHERPA[2.2.1]~\cite{Bothmann:2019yzt}                        & \SHERPA[2.2.1]~\cite{Schumann:2007mg}         & NNLO~\cite{Anastasiou:2003ds}                 & Dedicated     & \NNPDF[3.0nnlo]~\cite{Ball:2014uwa} & \NNPDF[3.0nnlo]~\cite{Ball:2014uwa} \\
Diboson                 &  \SHERPA[2.2.1],\,2.2.2~\cite{Bothmann:2019yzt}      &       \SHERPA[2.2.1],\,2.2.2~\cite{Gleisberg:2008fv,Schumann:2007mg}
& NLO~\cite{Hoeche:2011fd,Hoeche:2012yf,Catani:2001cc,Hoeche:2009rj}                  &  Dedicated  
&\NNPDF[3.0nnlo]~\cite{Ball:2014uwa}  & \NNPDF[3.0nnlo]~\cite{Ball:2014uwa} \\
 
$t\bar{t}$ 				& \POWHEGBOX[v2]~\cite{Frixione:2007nw,Nason:2004rx,Frixione:2007vw,Alioli:2010xd}       		& \PYTHIA[8.230]~\cite{Sjostrand:2014zea}  		& NNLO+NNLL~\cite{Beneke:2011mq,Cacciari:2011hy,Baernreuther:2012ws,Czakon:2012zr,Czakon:2012pz,Czakon:2013goa,Czakon:2011xx} 		& A14~\cite{ATL-PHYS-PUB-2014-021} & \NNPDF[3.0nlo]~\cite{Ball:2014uwa} & \NNPDF[2.3lo]~\cite{Ball:2012cx}\\
Single-top: \\
 
$~~~~Wt$  				& \POWHEGBOX[v2]~\cite{Re:2010bp,Nason:2004rx,Frixione:2007vw,Alioli:2010xd} 	& \PYTHIA[8.230]~\cite{Sjostrand:2014zea}	&NLO+NNLL~\cite{Kidonakis:2010ux,Kidonakis:2013zqa}		& A14~\cite{ATL-PHYS-PUB-2014-021} 			& \NNPDF[3.0nlo]~\cite{Ball:2014uwa}
& \NNPDF[2.3lo]~\cite{Ball:2012cx}
\\
~~~~$s$- and $t$-channels                       & \POWHEGBOX[v2]~\cite{Re:2010bp,Nason:2004rx,Frixione:2007vw,Alioli:2010xd}    & \PYTHIA[8.230]~\cite{Sjostrand:2014zea}        &NLO~\cite{Aliev:2010zk,Kant:2014oha}             & A14~\cite{ATL-PHYS-PUB-2014-021}                   & \NNPDF[3.0nlo]~\cite{Ball:2014uwa}
& \NNPDF[2.3lo]~\cite{Ball:2012cx}
\\
 
$t\bar{t}{+}H$ 	& \POWHEGBOX[v2]~\cite{Frixione:2007nw,Nason:2004rx,Frixione:2007vw,Alioli:2010xd,Hartanto:2015uka}      		& \PYTHIA[8.230]~\cite{Sjostrand:2014zea} 		& NLO	~\cite{Alwall:2014hca} & A14~\cite{ATL-PHYS-PUB-2014-021} & \NNPDF[3.0nlo]~\cite{Ball:2014uwa} & \NNPDF[2.3lo]~\cite{Ball:2012cx}\\
$t\bar{t}{+}V$     & \MGNLO[2.3.3]~\cite{Alwall:2014hca}      & \PYTHIA[8.210]~\cite{Sjostrand:2014zea}  & NLO~\cite{deFlorian:2016spz}      & A14~\cite{ATL-PHYS-PUB-2014-021} & \NNPDF[3.0nlo]~\cite{Ball:2014uwa} & \NNPDF[2.3lo]~\cite{Ball:2012cx} \\
$V H (\rightarrow\tau\tau)$           &\POWHEGBOX[v2]~\cite{Nason:2009ai,Nason:2004rx,Frixione:2007vw,Alioli:2010xd,Hamilton:2013fea,Hamilton:2015nsa}                    & \PYTHIA[8.212]~\cite{Sjostrand:2014zea}    &NNLO + NLO  \cite{Ciccolini:2003jy,Brein:2003wg,Brein:2011vx,Altenkamp:2012sx,Denner:2014cla,Brein:2012ne,Harlander:2014wda}              
&AZNLO~\cite{STDM-2012-23}      & \NNPDF[3.0nlo]~\cite{Ball:2014uwa}   & \PDFforLHC[15nnlo]~\cite{Butterworth:2015oua} \\

\bottomrule
 
\end{tabular}
 
\end{table}

For the simulated samples of $Z$ and $W$ bosons produced in association with jets (\Zjets and \Wjets) the \SHERPA[2.2.1]~\cite{Bothmann:2019yzt} generator was employed. Next-to-leading-order (NLO) matrix elements were used for up to two partons, while leading-order (LO) matrix elements were used for up to four partons. The Comix~\cite{Gleisberg:2008fv} and \OPENLOOPS~\cite{Buccioni:2019sur,Cascioli:2011va,Denner:2016kdg} libraries were utilised
for their computation. \SHERPA parton showers~\cite{Schumann:2007mg} were employed, with the \MEPSatNLO prescription~\cite{Hoeche:2011fd,Hoeche:2012yf,Catani:2001cc,Hoeche:2009rj} used to match the parton showers to the matrix elements. The \NNPDF[3.0nnlo] set of parton distribution functions (PDFs)~\cite{Ball:2014uwa} was utilised, and the samples were normalised to next-to-next-to-leading-order (\NNLO) cross-section predictions~\cite{Anastasiou:2003ds}.
 
Samples of diboson and triboson production were simulated with
the \SHERPA[2.2.1] or 2.2.2~\cite{Bothmann:2019yzt} generator, depending on the process. The matrix elements were matched and merged with the \SHERPA parton showers based on Catani--Seymour dipole factorisation~\cite{Gleisberg:2008fv,Schumann:2007mg} using the \MEPSatNLO prescription~\cite{Hoeche:2011fd,Hoeche:2012yf,Catani:2001cc,Hoeche:2009rj}. The \OPENLOOPS library~\cite{Buccioni:2019sur,Cascioli:2011va,Denner:2016kdg} was used to provide the virtual QCD corrections to the matrix elements where relevant. The samples were generated with the \NNPDF[3.0nnlo] PDF set~\cite{Ball:2014uwa}.
 
All \SHERPA -produced samples used a dedicated set of tuned parton-shower parameters developed by the \SHERPA authors~\cite{Gleisberg:2008fv,Schumann:2007mg}.

Higgs boson production and decay were simulated using the \POWHEGBOX[v2]~\cite{Nason:2009ai,Alioli:2010xd,Nason:2004rx,Frixione:2007vw,Hamilton:2013fea,Hamilton:2015nsa} generator interfaced with \PYTHIA[8]~\cite{Sjostrand:2014zea} for parton showering and non-perturbative effects. The \PDFforLHC[15nlo] PDF set~\cite{Butterworth:2015oua} and the \AZNLO tune~\cite{STDM-2012-23} of \PYTHIA[8] were used. The gluon--gluon fusion prediction from \POWHEG was normalised to the next-to-next-to-next-to-leading-order cross-section in QCD plus electroweak corrections at NLO~\cite{deFlorian:2016spz,Anastasiou:2016cez,Anastasiou:2015ema,Dulat:2018rbf,Harlander:2009mq,Harlander:2009bw,Harlander:2009my,Pak:2009dg,Actis:2008ug,Actis:2008ts,Bonetti:2018ukf,Bonetti:2018ukf}. The vector-boson fusion prediction was normalised to an approximate-NNLO QCD cross-section with NLO electroweak corrections~\cite{Ciccolini:2007jr,Ciccolini:2007ec,Bolzoni:2010xr}. The Monte Carlo
prediction of Higgs boson production in association with a vector boson was normalised to cross-sections calculated at NNLO in QCD with NLO electroweak corrections for \(q\bar{q}/qg \to VH\) and at NLO
and next-to-leading-logarithm accuracy in QCD for \(gg \to ZH\)~\cite{Ciccolini:2003jy,Brein:2003wg,Brein:2011vx,Altenkamp:2012sx,Denner:2014cla,Brein:2012ne,Harlander:2014wda}.

The production of top--antitop quark (\ttbar) events was simulated with the \POWHEGBOX[v2]~\cite{Frixione:2007nw,Nason:2004rx,Frixione:2007vw,Alioli:2010xd}
generator at the \NLO accuracy. The \NNPDF[3.0nlo]~\cite{Ball:2014uwa} PDF set was used, \mtop was set to $172.5$~\GeV, and the \hdamp parameter
was fixed at 1.5\,\mtop~\cite{ATL-PHYS-PUB-2016-020}. \PYTHIA[8.230]~\cite{Sjostrand:2014zea} with the A14 tune~\cite{ATL-PHYS-PUB-2014-021} was used to model the parton shower, hadronisation, and underlying event. The \NNPDF[2.3lo] ~\cite{Ball:2012cx} PDF set was used.
 
Single-top-quark processes were similarly modelled with the \POWHEGBOX[v2]~\cite{Re:2010bp,Nason:2004rx,Frixione:2007vw,Alioli:2010xd} generator using the \NNPDF[3.0nlo] set of PDFs~\cite{Ball:2014uwa}. The associated production of a top quark with a $W$ boson ($Wt$) and $s$-channel production utilised the five-flavour scheme, while the $t$-channel production used the four-flavour scheme. For $Wt$ processes the diagram removal scheme~\cite{Frixione:2008yi} was used to remove interference with the \ttbar samples. The events were then interfaced to \PYTHIA[8.230]~\cite{Sjostrand:2014zea}, which used the A14 tune~\cite{ATL-PHYS-PUB-2014-021} and the \NNPDF[2.3lo] set of PDFs~\cite{Ball:2012cx}. The same configuration also applies to the production of \ttH events.
 
The inclusive cross-sections for the previously described top-quark process samples were corrected to include higher-order effects. For \ttbar, the cross-section was calculated to \NNLO in perturbative QCD with the resummation of next-to-next-to-leading logarithmic (NNLL) soft-gluon terms using the \TOPpp program~\cite{Beneke:2011mq,Cacciari:2011hy,Baernreuther:2012ws,Czakon:2012zr,Czakon:2012pz,Czakon:2013goa,Czakon:2011xx}. For $Wt$ events, the corrections were to NLO+NNLL accuracy ~\cite{Kidonakis:2010ux,Kidonakis:2013zqa}, while $s$- and $t$-channel production was corrected to \NLO accuracy~\cite{Aliev:2010zk,Kant:2014oha}. The \ttH cross-section was corrected to \NLO QCD+electroweak accuracy~\cite{deFlorian:2016spz}.
 
The \ttV\ ($V= Z, W$) samples were produced with
the \MGNLO[2.3.3]~\cite{Alwall:2014hca} generator at \NLO accuracy
using the \NNPDF[3.0nlo]~\cite{Ball:2014uwa} PDF set. The matrix
elements were interfaced
with \PYTHIA[8.210]~\cite{Sjostrand:2014zea}. The A14
tune~\cite{ATL-PHYS-PUB-2014-021} and
the \NNPDF[2.3lo]~\cite{Ball:2014uwa} PDF set were used.
 
Decays of bottom and charm hadrons for top-quark processes, including \ttH, were performed by \EVTGEN[1.6.0]~\cite{Lange:2001uf}.

For the \TwoHDMa samples, the matrix elements were generated
with \MGNLO[2.6.5]~\cite{Alwall:2014hca} at
LO. \PYTHIA[8.240]~\cite{Sjostrand:2014zea} with the A14
tune~\cite{ATL-PHYS-PUB-2014-021} was used to
model the parton shower, hadronisation and underlying event. The CKKW-L
merging procedure~\cite{Lonnblad:2001iq,Lonnblad:2011xx} was employed
to match the matrix element to the parton showers. Signal Monte Carlo
samples were generated with the \NNPDF[3.0nlo] PDF
set~\cite{Ball:2012cx} and processed with a fast simulation that
relies on a parameterisation of the calorimeter
response~\cite{SOFT-2010-01}. Both the gluon--gluon fusion and $b\bar{b}$
annihilation signal production processes were considered.
 
The \TwoHDMa samples were generated with different choices of signal
model parameters, varying two parameters at a time. In the first
parameter scan, the pseudoscalar singlet mass $m_a$ was varied in the range
$[100,400]$~\GeV\ while the mass of the pseudoscalar Higgs boson $m_A$ was varied in the range $[300,1400]$~\GeV. The requirements $\tan \beta = 1$ and $\sin\Theta = 0.35$ were applied. In the second parameter scan, $m_A$ was varied in the range $[500,1300]$~\GeV\ with $\tan \beta$ taking values from $0.3$ to $20$. In this scan, constant values of $m_a = 250$~\GeV\ and $\sin\Theta = 0.7$ were used. The particular parameter choices are based on the recommendations from the LHC DM Working Group \cite{Abe:2644694}.

For all simulated samples, the effect of multiple interactions in the same and neighbouring bunch crossings (\pileup) was modelled by overlaying each simulated hard-scattering event with inelastic proton--proton events generated with \PYTHIA[8.186]~\cite{Sjostrand:2007gs} using the \NNPDF[2.3lo] set of PDFs~\cite{Ball:2012cx} and the A3 tune~\cite{ATL-PHYS-PUB-2016-017}.
 
To compensate for differences between data and simulation, dedicated per-object correction factors derived from data are applied to jets, $b$-jets, $\tau$-leptons, electrons, and muons. Additional per-event correction factors are applied to account for differences in trigger efficiency between data and simulation.


\section{Event reconstruction}
\label{sec:reconstruction}


An event is considered to have a primary vertex if at least two
ID tracks with $\pt > 500$~\MeV\ can be associated with
it~\cite{ATL-PHYS-PUB-2015-026}. If there are several such vertices, the candidate
with the highest sum of the squared transverse momenta of associated
tracks, $\Sigma \pt ^2$, is considered as the hard-scattering vertex.

Jet candidates are reconstructed from particle-flow objects \cite{PERF-2015-09} using the anti-$k_t$ algorithm \cite{Cacciari:2008gp} with a distance parameter of $R = 0.4$.
First, the calorimeter cells are grouped into topological clusters, with the energy measured at the electromagnetic scale to which all ATLAS calorimeters are calibrated. The energy deposited by the charged particles is then subtracted from the clusters and replaced by the momenta of matching tracks. The jet energy scale (JES)~\cite{JETM-2018-05} calibration is applied to the jet candidates, restoring the energy to that measured at particle level.
The jets are required to have a transverse momentum larger than $20$~\GeV\ and $|\eta| < 2.5$. In addition, jets with $\pt < 60~\GeV$ are required to pass the \textit{tight} working point of the jet-vertex-tagger (JVT) algorithm \cite{PERF-2014-03} to suppress background from \pileup interactions.
 
The analysis aims for a selection orthogonal to that of the mono-Higgs search with $b$-jets \cite{EXOT-2018-46}, so robust $b$-tagging with the same selection efficiency is important.
Jets containing $b$-hadrons are identified as $b$-jets using the
classification algorithm DL1r \cite{ATL-PHYS-PUB-2017-013} based on
deep-learning techniques. The DL1r tagger uses secondary-vertex
displacement information and kinematic properties of tracks to select
$b$-jets and reject jets originating from charm or light quarks. The
$b$-tagging working point used for the analysis has an efficiency of
$77\%$, as measured in an inclusive \ttbar sample, while providing rejection factors of $4.9$, $14$ and $130$ against $c$-jets, $\tau$-jets, and light-flavour jets, respectively \cite{FTAG-2018-01}.

Electrons are reconstructed from energy deposits in the electromagnetic calorimeter that are matched to an ID track. Electron candidates are required to satisfy the \textit{loose} identification criteria \cite{PERF-2017-01,EGAM-2018-01}, and to have $\pt > 10~\GeV$ and $|\eta| < 2.47$, providing selection efficiency of $93\%$. Additionally, the longitudinal impact parameter $z_0$ has to satisfy $|z_0 \sin \theta| < 0.5$~mm to discard electron candidates not associated with the primary vertex. The objects meeting these selection criteria are called baseline electrons and participate in the computation of the missing transverse momentum and in the overlap-removal procedure, as described later.
Baseline electrons that also satisfy $\pt > 25$~\GeV, fulfil
the \textit{loose} isolation criteria and the \textit{tight}
identification criteria, and satisfy the condition $|d_{0}| / \sigma(d_{0}) <
5$ on the transverse impact parameter $d_{0}$ and its uncertainty
$\sigma(d_{0})$ are henceforth simply referred to as electrons.

Muons are reconstructed from tracks in the muon spectrometer that are matched to tracks in the ID. Muon candidates are required to satisfy the \textit{medium} identification criteria~\cite{PERF-2015-10,MUON-2018-03} and to have $\pt > 10$~\GeV\ and $|\eta| < 2.7$. This working point has an efficiency of $97\%$. Additionally, the longitudinal impact parameter has to satisfy $|z_0 \sin \theta| < 0.5$~mm. The objects meeting these selection criteria are called baseline muons and participate in the computation of the missing transverse momentum and in the overlap-removal procedure.
Baseline muons that also satisfy $\pt > 25$~\GeV, fulfil
the \textit{loose} isolation criteria, and satisfy the condition $|d_{0}|
/ \sigma(d_{0}) < 3$ are henceforth simply
referred to as muons.
 
Reconstruction of the visible part of a hadronically decaying
$\tau$-lepton~\cite{ATL-PHYS-PUB-2015-045} is seeded by anti-$k_t$
jets with a distance parameter $R = 0.4$. The seed jet is built from
topological clusters, calibrated with a hadronic weighting scheme
\cite{Barillari:1112035}, and is required to have $\pt > 10$~\GeV\ and
$|\eta| < 2.5$. The $\tauhadvis$ candidate is then built from
tracks and clusters within $\Delta R = 0.2$ of the seed jet's axis. Only
candidates with one or three charged tracks with a sum of $\pm
1$ are considered. $\tauhadvis$ candidates are required to have $\pt >
20$~\GeV\ and $|\eta| < 2.5$. Candidates in the transition region
between the barrel and endcap calorimeters, $1.37 < |\eta| < 1.52$,
are excluded. The $\tauhadvis$ object's energy is calibrated with a boosted
regression tree using information from the calorimeter and the particle-flow
reconstruction, as well as the number of \pileup interactions~\cite{ATLAS-CONF-2017-029}, as input.
To suppress electron background, a boosted decision tree is used to reject events with electrons misidentified as \tauhadvis. It is trained using tracking detector and calorimeter information as well as variables related to the ratio of the energy deposited in the calorimeter and the visible momentum of the reconstructed tracks.

To further discriminate hadronically decaying $\tau$-leptons from background jets, a
recurrent neural network (RNN) algorithm is used \cite{ATL-PHYS-PUB-2019-033}. The algorithm uses tracking and calorimeter measurement information, as well as individual track and cluster information, as input. The analysis makes use of three different ways to define $\tau$-lepton objects using the RNN algorithm. The first type of object (baseline $\tau$-lepton) is required to pass the \textit{very loose} identification working point. These objects are used in the overlap-removal procedure. The baseline $\tau$-leptons that also pass the \textit{medium} identification working point are simply referred to as $\tau$-leptons. The last type of object considered by the analysis consists of $\tau$-lepton candidates that do not pass the \textit{medium} identification working point. Such objects are referred to as anti-ID $\tau$-leptons and are used by the analysis to estimate the contribution of fake $\tau$-leptons.

The missing transverse momentum \met\ is a measure of the transverse-momentum imbalance in the detector and is computed as the magnitude of the missing transverse momentum vector $\vecmet$~\cite{ATLAS-CONF-2018-023}. The latter is calculated as the negative vector sum of the transverse momenta of electrons, muons, $\tau$-leptons and jets. Tracks that are associated with the primary vertex but not with any reconstructed object are also included in the calculation as the so-called \enquote{soft term}.
 
The reconstructed objects are not exclusive, e.g.~a single energy deposit in the calorimeter can be used in the reconstruction of several final-state objects. In order to resolve this ambiguity, an overlap-removal procedure is applied~\cite{HIGG-2019-09} so that at most one reconstructed object is associated with a detector signal:
\begin{itemize}
\item If two baseline electrons share a track, only the baseline electron with the higher transverse momentum is considered.
\item A baseline $\tau$-lepton is discarded if it is closer than $\Delta R = 0.2$ to a baseline electron or a muon.
\item If a baseline muon and a baseline electron share an ID track the baseline electron is discarded. The only exception is when the baseline muon is calorimeter-tagged (i.e. a track in the ID matched to an energy deposit in the calorimeter compatible with a minimum-ionizing particle~\cite{PERF-2015-10,MUON-2018-03}), then the baseline muon is discarded instead.
\item If a jet is closer than $\Delta R = 0.2$ to a baseline electron or muon, it is discarded. The only exception is if the jet has three or more associated tracks, in which case it is kept instead of the baseline muon. This suppresses FSR and bremsstrahlung signatures that can be reconstructed as jets with low number of tracks.
\item If any of the remaining baseline electrons or muons are closer than $\Delta R = 0.4$ to the remaining jets, they are discarded.
\item If any of the remaining jets is closer than $\Delta R = 0.2$ to any of the remaining baseline $\tau$-leptons, the jet is discarded.
\end{itemize}
 
For anti-ID $\tau$-lepton candidates the overlap-removal procedure is performed in the same way, replacing the baseline $\tau$-lepton identification criterion with that for anti-ID $\tau$-leptons. However, two additional corrections are needed for the overlap-removal procedure compared to the one applied to baseline $\tau$-leptons. First, the jets that do not pass the \textit{tight} JVT working point selection requirements are given a priority over anti-ID $\tau$-leptons not meeting the \textit{very loose} identification criterion. This is done so that such jets are not masked by being treated as anti-ID $\tau$-lepton candidates and can be vetoed successfully. The second correction is to give the same priority to $b$-jets. Otherwise, anti-ID $\tau$-leptons can mask $b$-jets, whereas the analysis relies on an accurate counting of the number of $b$-jets.


\section{Event selection}
\label{sec:selection}

 
Data and simulated events are categorised in regions based on the properties of the reconstructed objects.
The analysis uses three types of regions -- control, validation, and signal regions. Signal regions (SRs) are regions where the sensitivity to the BSM signal model is maximised.
Control regions (CRs) are regions enriched in particular background processes and are kinematically close to the SRs.
A background-only fit of the simulated background processes to the data is performed in CRs using the \histfitter framework \cite{Baak:2014wma}.
In this fit the normalisation of the dominant background processes is determined.
The fitted normalisations are then propagated to the validation regions (VRs) and SRs. 
VRs are defined so that they lie between the SRs and CRs and serve to validate the extrapolation of background normalisation from the CRs to the SRs. The results of the background-only fit are discussed further in Section \ref{sec:result}.

Events used by the analysis are required to have a primary vertex.
The events are also required to pass a jet
quality criterion to reject events containing jets not
originating from proton--proton collisions~\cite{ATLAS-CONF-2015-029}, but from beam-induced backgrounds or cosmic-ray showers.

Events are required to have exactly two $\tau$-lepton objects that geometrically match the objects activating the di-$\tauhadvis$ part of the triggers.
Both the \met\ and the di-$\tauhadvis$ parts
of the trigger evolved during \RunTwo to implement more efficient
algorithms and adapt them to the data-taking conditions. While the HLT \met trigger requirement was kept constant at $50$~\GeV, the level-1
threshold was increased from $35$ to $40$~\GeV\ in
2018. The HLT threshold for the leading-$\tauhadvis$ $\pt$ was
increased from $35$~\GeV\ to $60$~\GeV\ in 2018, while the
sub-leading-$\tauhadvis$ $\pt$ threshold remained constant at
$25$~\GeV. The HLT $\tauhadvis$ identification algorithms were improved during \RunTwo, resulting in increased rejection power.
The leading $\tau$-lepton is required to have $\pt > 40(65)$~\GeV\ when matched to the HLT $\tauhadvis$ object passing the trigger's $35(60)$~\GeV\ threshold. The sub-leading $\tau$-lepton is required to have $\pt > 30$~\GeV. The events also have to satisfy the $\met > 150$~\GeV\ requirement at which the trigger is operating at maximum efficiency.
 
Events containing an electron or a muon are vetoed.
Events are required to have at most one $b$-jet.
The common analysis preselection is summarised in Table~\ref{tab:plateau_presel}, defining the analysis phase space. In the following, the analysis objects in an event are assumed to be ordered by decreasing transverse momentum with indices \enquote{1} and \enquote{2} referring to the leading and sub-leading objects respectively.

\begin{table}[t]
\caption{
Summary of the common analysis preselection. The upper part contains requirements related to optimising the trigger efficiency while the lower part contains analysis signature-related requirements.}
 
\label{tab:plateau_presel}
\centering
\begin{tabular}{lcc}
\toprule
Observable & \multicolumn{2}{c}{Selection} \\
\midrule
Year	  &  2015--2017 & 2018 \\
\midrule
$\pt^{\tau_1}$ & $> 40$~\GeV & $> 65$~\GeV \\
$\pt^{\tau_2}$ & \multicolumn{2}{c}{\phantom{0}$> 30$~\GeV } \\
$\met$ & \multicolumn{2}{c}{$ > 150$~\GeV} \\
$N_{\tau}$ &  \multicolumn{2}{c}{2} \\
\midrule
$N_{e}$ & \multicolumn{2}{c}{0} \\
$N_{\mu}$ & \multicolumn{2}{c}{0} \\
$N_{b\textrm{-jet}}$ & \multicolumn{2}{c}{$\leq 1$\phantom{$\leq$~}}\\
\bottomrule
\end{tabular}
\end{table}

In order to improve the sensitivity to the \TwoHDMa signal models
considered, two SRs are defined. This is done in
two steps. First, a set of common SR preselection requirements is applied to
suppress the dominant SM processes. After the common preselection, two non-orthogonal SRs are designed to target signal model parameter configurations with high and low masses of the heavy pseudoscalar Higgs boson $A$, as these yield signal events with different kinematic properties.
The combination of the results from the two SRs is described in Section~\ref{sec:result}.
 
The common SR preselection includes a veto on events with $b$-jets and
a requirement that the two $\tau$-leptons have opposite electric charges.
The invariant mass of the two \tauhadvis, or \enquote{visible invariant mass}
$m_\text{vis}(\tau_1,\tau_2)$, is constrained to be between
$40$~\GeV\ and $125$~\GeV\ to ensure compatibility with a Higgs boson decay.
The angular distance between the two $\tau$-leptons is required to satisfy $\Delta R < 2$,
suppressing background events in which the two $\tau$-leptons do not originate from a resonant decay (such as \ttbar\ and \Wjets). Two kinematic variables combining \met\ and the $\tau$-leptons' \pt\ are used to suppress \Zjets events. The event-level $m_\text{T}^\text{tot}$ is defined as
\begin{equation}  \nonumber
m_\text{T}^\text{tot} = \sqrt{(\pt^{\tau_{1}} + \pt^{\tau_{2}} + \pt^\text{miss})^2 - (p_x^{\tau_{1}} + p_x^{\tau_{2}} + p_x^\text{miss})^2 - (p_y^{\tau_{1}} + p_y^{\tau_{2}} + p_y^\text{miss})^2}\ ,
\end{equation}
where $p_x^\text{miss}$ and $p_y^\text{miss}$ correspond to the $x$ and $y$ components of the missing transverse momentum vector. An event is required to have $m_\text{T}^\text{tot} > 50$~\GeV\ to pass the common SR preselection, while $m_\text{T}^\text{tot} < 50$~\GeV\ selection is used for trigger efficiency studies. The transverse mass of a $\tau$-lepton is a per-object variable defined as
\begin{equation} \nonumber
m_\text{T}^{\tau_{i}} = \sqrt{2\pt^{\tau_{i}} \met(1 - \cos \Delta \phi(\tau_{i},\pt^\text{miss}))} \ .
\end{equation}
The sum $m_\text{T}^{\tau_{1}} + m_\text{T}^{\tau_{2}}$ is required
to be larger than $100$~\GeV\ to suppress events in which \met\ is
collinear with the two-$\tau$-lepton system, as this is typical for $Z$
boson event topologies. A summary of the common SR preselection is presented in the upper part of Table~\ref{tab:signal_selection}.
 
A \enquote{Low$_{m_A}$} SR is defined on top of the common SR preselection to target signal models with $m_A \leq 800~\GeV$. The angular distance between the two $\tau$-leptons is tightened to be $0.6 < \Delta R(\tau_1, \tau_2) < 1.9$ and the visible invariant mass of the two $\tau$-leptons is required to be larger than $75$~\GeV. The transverse masses of the leading and sub-leading $\tau$-leptons are required to be $m_\text{T}^{\tau_1} >50$~\GeV\ and $m_\text{T}^{\tau_2} >25$~\GeV. The SR is divided into four $m_\text{T}^{\tau_{1}} + m_\text{T}^{\tau_{2}}$ bins: $<250$~\GeV, $[250,400]$~\GeV, $[400,550]$~\GeV, and $>550$~\GeV.
 
A \enquote{High$_{m_A}$} SR is constructed to improve the sensitivity to models with high $m_A$ masses, leading to signal events with higher \met\ and boosted Higgs bosons. In addition to the common SR preselection, this SR requires $m_\text{T}^\text{tot} > 400$~\GeV\ and  $m_\text{T}^{\tau_{1}} + m_\text{T}^{\tau_{2}} > 400$~\GeV. Similarly to the Low$_{m_A}$ SR, the events in the High$_{m_A}$ SR are separated in two $m_\text{T}^{\tau_{1}} + m_\text{T}^{\tau_{2}}$ bins, above and below $750~\GeV$.

\begin{figure}[t]
\centering
\subfloat[]{\includegraphics[width= 0.45\textwidth]{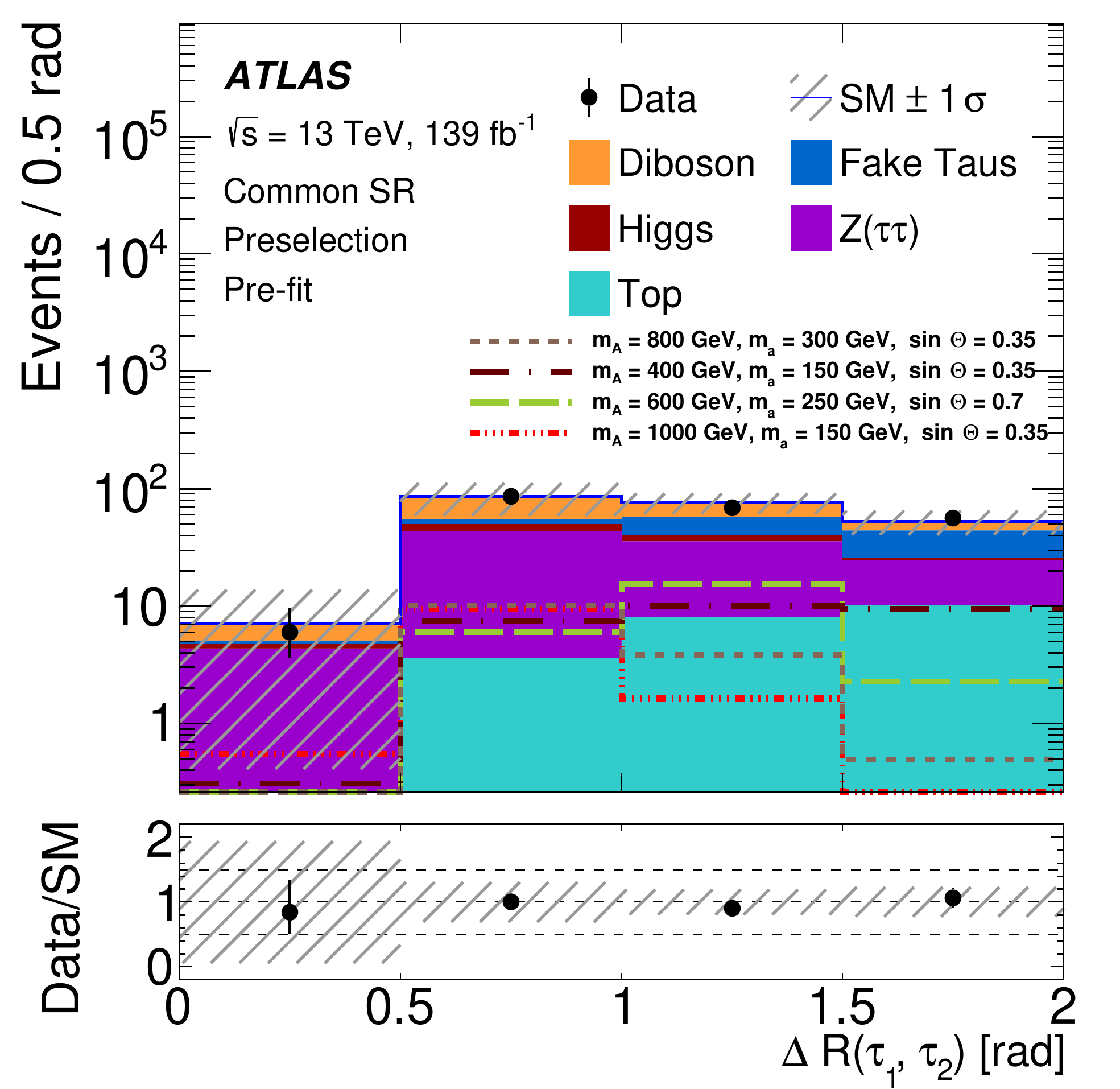}\label{fig:lowmASRcomb_ditau_DR}}
\hfil
\subfloat[]{\includegraphics[width= 0.45\textwidth]{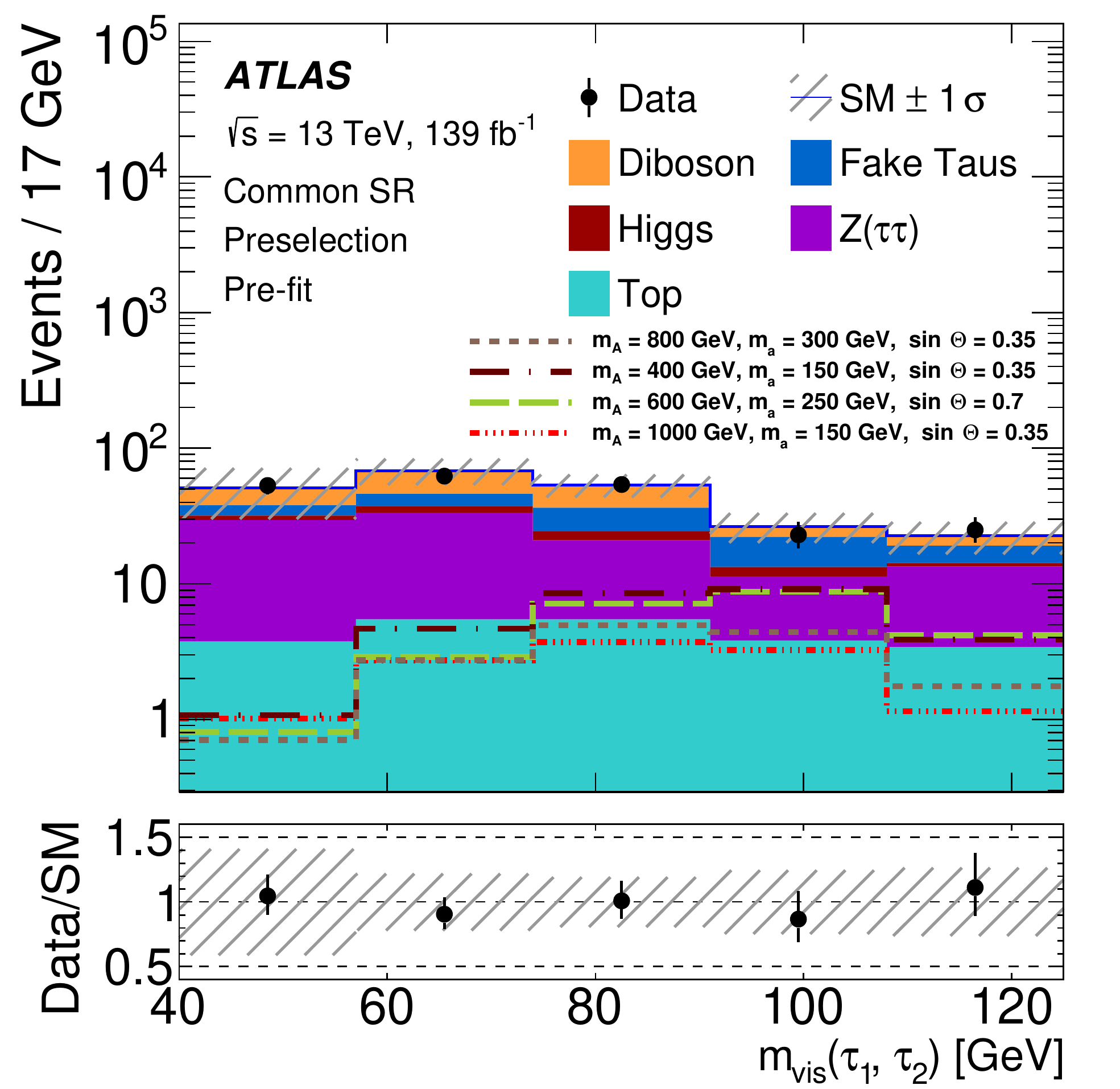}\label{fig:lowmASRcomb_ditau_mass}}\\
\subfloat[]{\includegraphics[width= 0.45\textwidth]{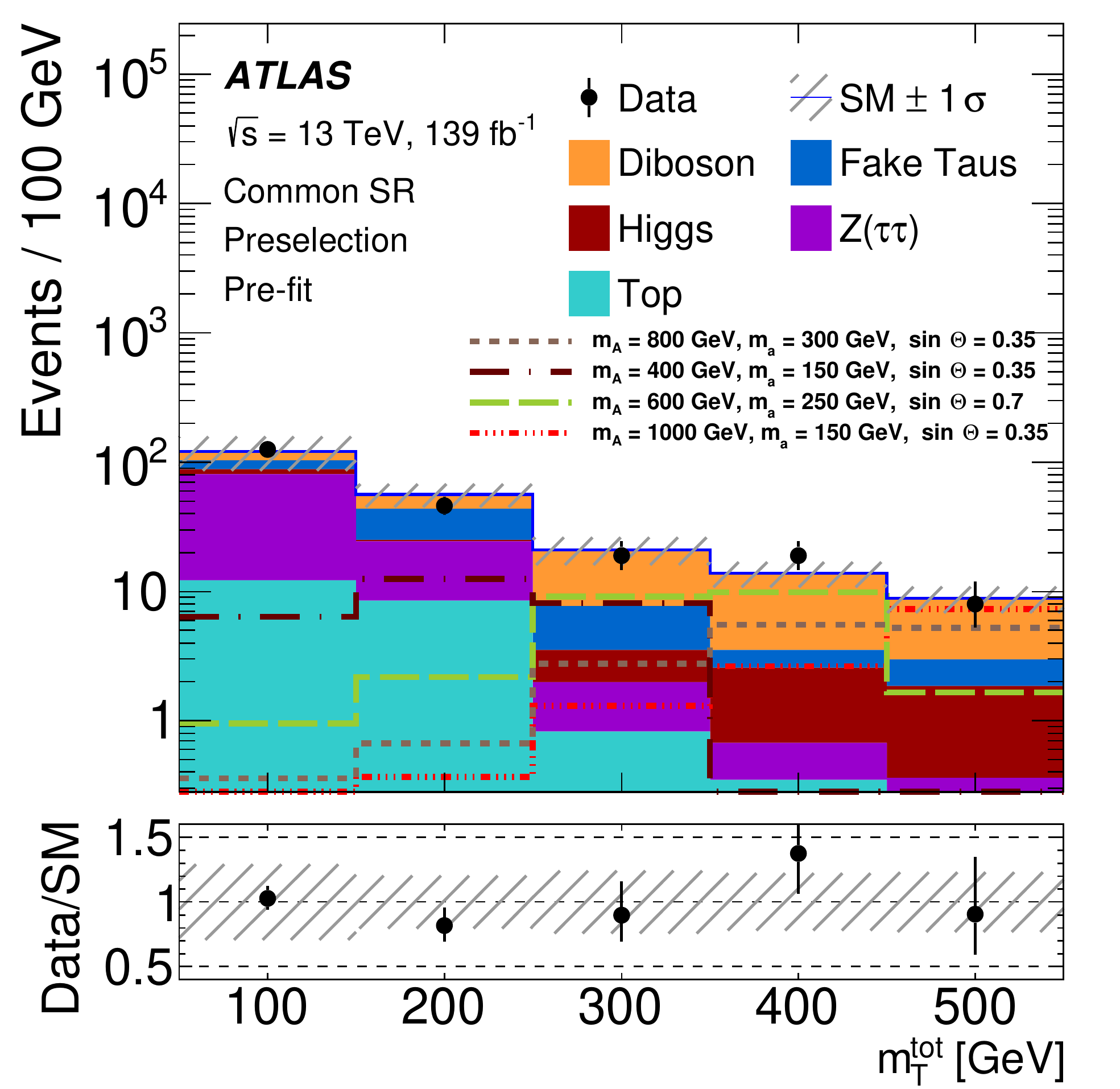}\label{fig:lowmASRcomb_mTtot}}
\hfil
\subfloat[]{\includegraphics[width= 0.45\textwidth]{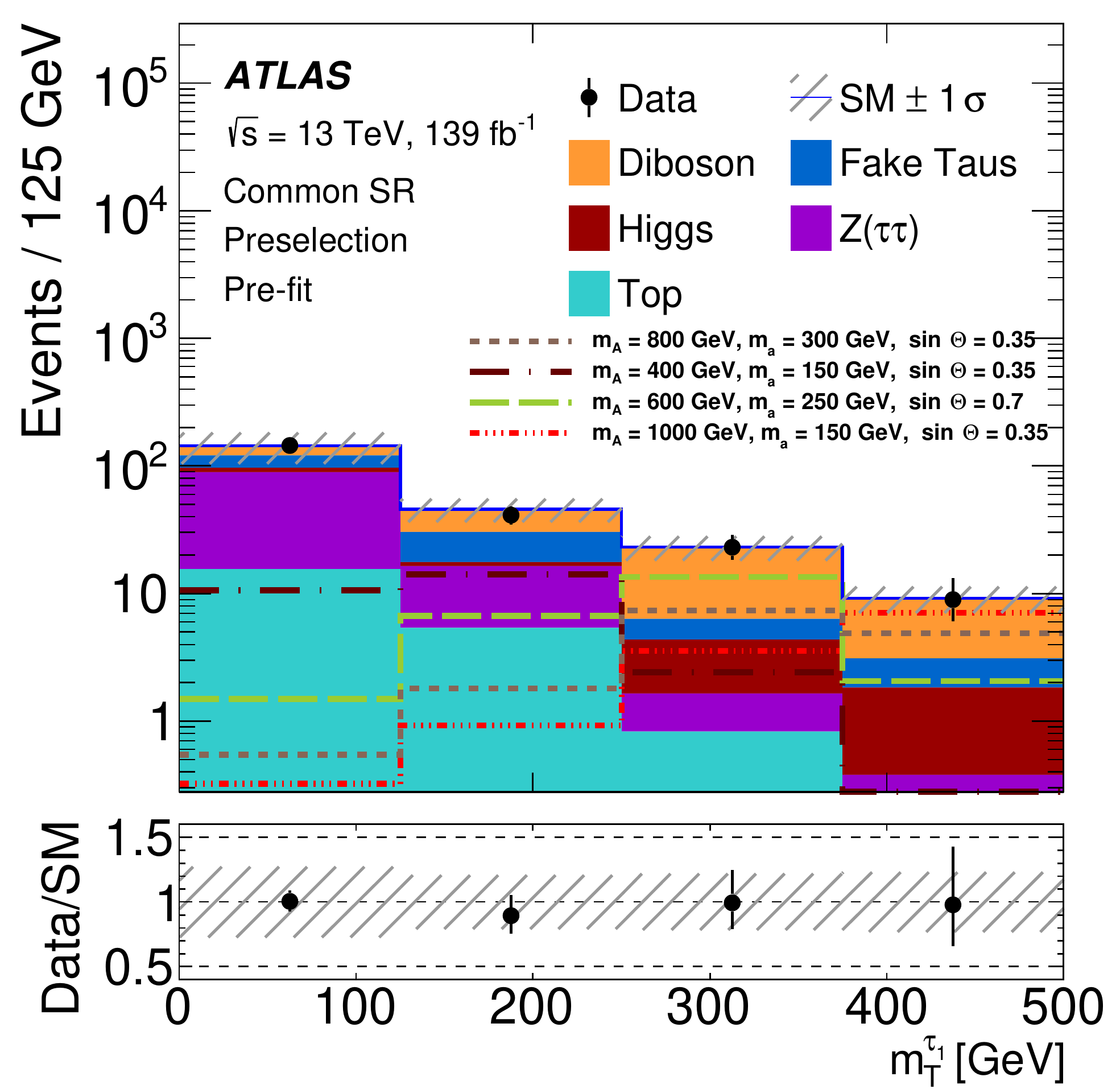}\label{fig:lowmASRcomb_tau_1_mtMet}}
\caption{Kinematic distributions of some of the variables used to construct the SRs, shown in the common SR preselection region:
\protect\subref{fig:lowmASRcomb_ditau_DR} $\Delta R(\tau_1, \tau_2)$,
\protect\subref{fig:lowmASRcomb_ditau_mass} $m_\text{vis}(\tau_1,\tau_2)$,
\protect\subref{fig:lowmASRcomb_mTtot} $m_\text{T}^\text{tot}$,
\protect\subref{fig:lowmASRcomb_tau_1_mtMet}      $ m_\text{T}^{\tau_1}$.
Predictions for four benchmark parameter configurations of the \TwoHDMa model are also shown for the parameter values given in the legend. The value $\tan(\beta) = 1$ is assumed for all parameter configurations.
\BandIsStatSyst
\RightBinOverflow
\BottomPanelDataSM
}
\label{fig:common_SR}
\end{figure}

A summary of the High$_{m_A}$ and Low$_{m_A}$ SRs' requirements is presented in the lower part of Table~\ref{tab:signal_selection}. Examples of kinematic distributions of the important variables ($\Delta R(\tau_1, \tau_2)$, $m_\text{vis}(\tau_1,\tau_2)$, $m_\text{T}^\text{tot}$, $m_\text{T}^{\tau_1}$) after applying the common SR preselection are shown in Figure~\ref{fig:common_SR}. The way the SM background estimations are performed is described in Section~\ref{sec:background}. Predictions from four benchmark parameter configurations of the \TwoHDMa model normalised to the theoretical cross-section and the integrated luminosity of the data are also shown in Figure~\ref{fig:common_SR}.
 
The combined acceptance and efficiency of the Low$_{m_A}$ SR bins varies from $0.0013$ to $3 \cdot 10^{-6}$ over all of the parameter configurations of the \TwoHDMa model considered by the analysis for gluon--gluon fusion production. For the High$_{m_A}$ SR bins the combined acceptance and efficiency varies from $0.0023$ to $0$ because the High$_{m_A}$ SR is only designed to be sensitive to a subset of all the parameter configurations.

\begin{table}[t]
\caption{Summary of the selection requirements used to define the signal regions. The upper part of the table describes the common SR preselection while the lower part of the table describes specific selections to define the Low$_{m_A}$ and High$_{m_A}$ SRs. These are applied in addition to the common analysis preselection described in Table~\ref{tab:plateau_presel}.}
\label{tab:signal_selection}
\centering
\begin{tabular}{lcc}
\toprule
&  \multicolumn{2}{c}{Common SR Preselection} \\
\midrule
$\Delta R(\tau_1, \tau_2)$ & \multicolumn{2}{c}{$<2$} \\
$m_\text{T}^\text{tot}$ & \multicolumn{2}{c}{\phantom{0}$>50$~\GeV} \\
$m_\text{vis}(\tau_1,\tau_2)$ & \multicolumn{2}{c}{$ \in [40,125]$~\GeV} \\
$m_\text{T}^{\tau_1} + m_\text{T}^{\tau_2}$ & \multicolumn{2}{c}{$>100$~\GeV} \\
Charge($\tau_1, \tau_2$) & \multicolumn{2}{c}{$q(\tau_1) \times q(\tau_2) = -1$} \\
$N_{b\textrm{-jet}}$ & \multicolumn{2}{c}{0} \\
\midrule
& Low$_{m_A}$ SR & High$_{m_A}$ SR \\
\midrule
$\Delta R(\tau_1, \tau_2)$ &  $\in[0.6,1.9]$ & $<2$ \\
$m_\text{T}^\text{tot}$ & - & $>400$~\GeV \\
$m_\text{T}^{\tau_1}$ & $>50$~\GeV  & - \\
$m_\text{T}^{\tau_2}$ & $>25$~\GeV &  - \\
$m_\text{vis}(\tau_1,\tau_2)$ & $ >75$~\GeV &  $\in [40,125]$~\GeV \\
$m_\text{T}^{\tau_1} + m_\text{T}^{\tau_2}$ binning   & $[100,250,400,550,\infty]$~\GeV & $[400,750,\infty]$~\GeV \\
\bottomrule
\end{tabular}
\end{table}

\FloatBarrier
 
\section{Background estimation}
\label{sec:background}

Two sources of events are distinguished to perform the SM background estimation
-- those with both reconstructed $\tauhadvis$ objects originating from the
hadronic decays of prompt $\tau$-leptons (true $\tau$-leptons), and events with at least
one of the reconstructed $\tauhadvis$ objects being a non-$\tau$-lepton
object that satisfied the selection criteria, also referred to as a fake $\tau$-lepton.
Events from $Z$ boson and multiboson production, and the majority of \ttbar\ events, tend to have two true $\tau$-leptons in the analysis phase space.
Events from both $W$ boson and multijet production, as well as approximately $25\%$ of \ttbar\ events, are expected to have at least one fake $\tau$-lepton.

SM background processes are modelled using a combination of simulated events and data-driven methods.
Events with only true $\tau$-leptons are modelled using simulation normalised to the data in the dedicated CRs. In the following, when referring to the \ttbar\ background modelling, for example, it is only the events with two true $\tau$-leptons that are being addressed. Events with at least one fake $\tau$-lepton are estimated in a data-driven way. Such events are referred to as fake $\tau$-leptons or \enquote{Fake Taus} background regardless of the underlying physical process.
 
The normalisation of $Z$ boson production and various processes including top-quark production
(\ttbar\ production, single-top-quark production, \ttbar
production in association with $Z$/$W$) is fitted to the
data in dedicated control regions. 
Other SM background contributions are normalised according to the theoretical predictions.
For the derivation of the normalisation scale factors all top-quark-related processes are treated as one common \enquote{Top}
background.

The contribution of fake $\tau$-lepton events (i.e.~events with at
least one fake $\tau$-lepton) in the control, validation and signal
regions is estimated with the fake factor
method~\cite{HIGG-2017-07}. The assumption is that the
probability that a non-$\tau$-lepton object (such as a jet) meets the
$\tau$-lepton identification criteria depends purely on the
object's properties, such as its \pt, $\eta$, number of charged-particle tracks,
and origin (i.e.\ quark/gluon), and not on event-level variables such
as \met\ or number of $\tau$-leptons. 
The ratio of the number of objects that pass the \textit{medium} identification working point
requirement to the number of objects that do not, but instead pass the much looser anti-ID $\tau$-lepton selection, is called the fake factor. The fake factors were measured as a function of $\tau$-lepton's \pt, $|\eta|$, number of charged-particle tracks, and origin in regions enriched with fake $\tau$-leptons and extrapolated to the analysis phase space. In the analysis regions with the $\tau$-lepton selection requirement replaced by that for anti-ID $\tau$-leptons, the fake factors are applied to the anti-ID $\tau$-leptons to obtain a prediction of the fake $\tau$-lepton background.
 
The Fake Tau VR is used to validate the modelling of fake $\tau$-leptons. The events are required to have been recorded by the di-$\tau + \met$ trigger, to pass the common analysis preselection as described in Table~\ref{tab:plateau_presel}, and to have no $b$-jets. To increase the fake $\tau$-lepton purity of this VR, the two $\tau$-leptons are required to have the same charge. The two $\tau$-leptons are also required to have $\Delta R (\tau_1, \tau_2) < 2$ to remain kinematically close to the SRs. This selection ensures that the Fake Tau VR has a purity of over $85\%$.

Two CRs are defined by inverting or relaxing some of the common SR
requirements (see Table~\ref{tab:signal_selection}) such that they are
kinematically close to the SRs,
suppress a potential signal contribution, and are enriched in $Z$ boson and \ttbar\ backgrounds respectively.
Additionally, the analysis uses four VRs. Kinematically, the VRs are defined to be between the CRs and SRs and serve to validate the extrapolation of the modelling from the CRs to the SRs.
The definitions of CRs and VRs are summarised in Table \ref{tab:crvr_selection} and are further discussed in the following.
 
\begin{figure}[t]
\centering
\subfloat[$m_\text{vis}(\tau_1,\tau_2)$ in Top CR]{\includegraphics[width= 0.45\textwidth]{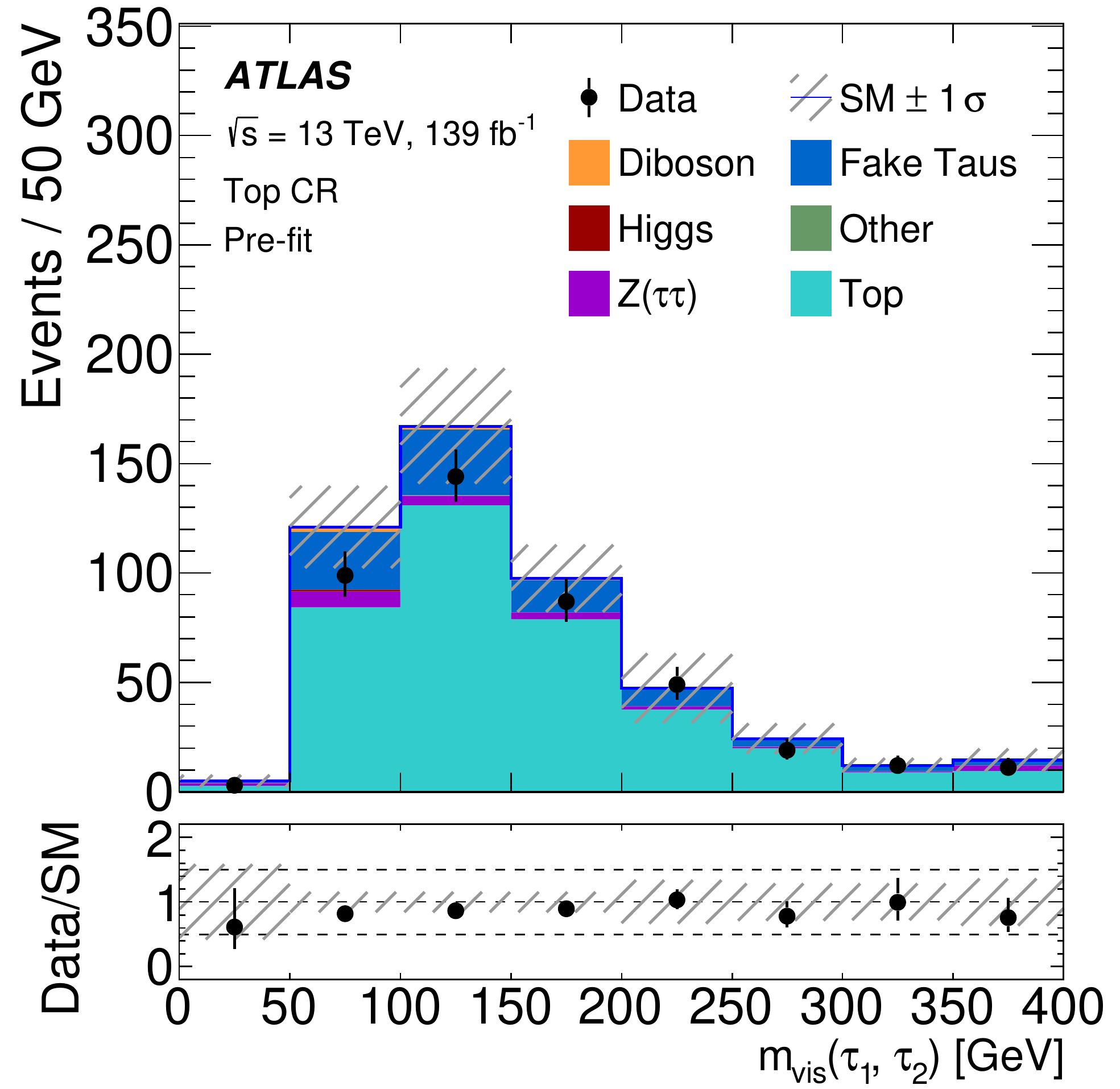}\label{fig:topCR_ditau_mass}}
\hfil
\subfloat[$ m_\text{T}^{\tau_2}$ in Top CR]{\includegraphics[width= 0.45\textwidth]{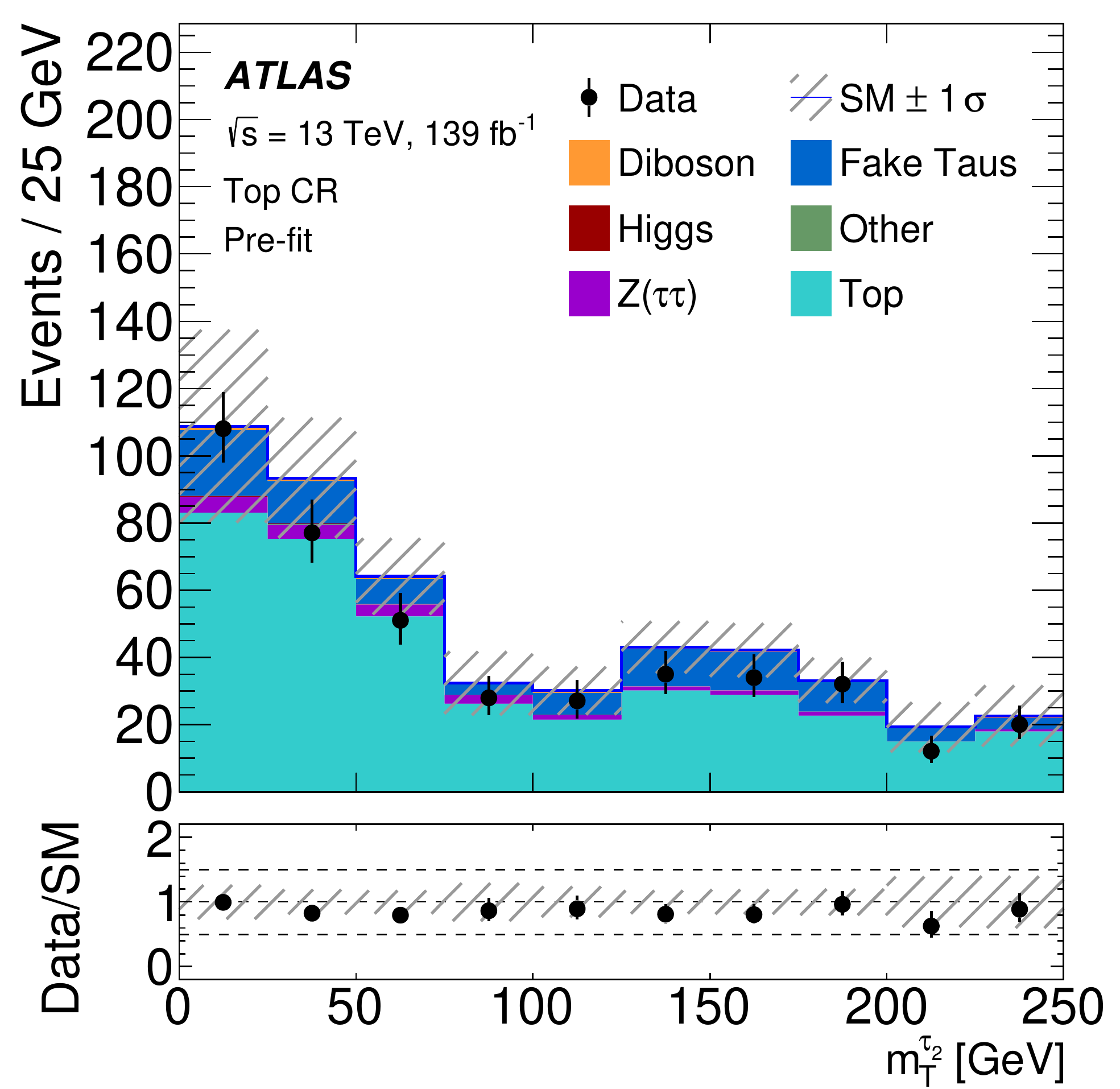}\label{fig:topCR_tau_2_mtMet}}\\
\subfloat[$\Delta R(\tau_1, \tau_2)$  in $Z$ CR]{\includegraphics[width= 0.45\textwidth]{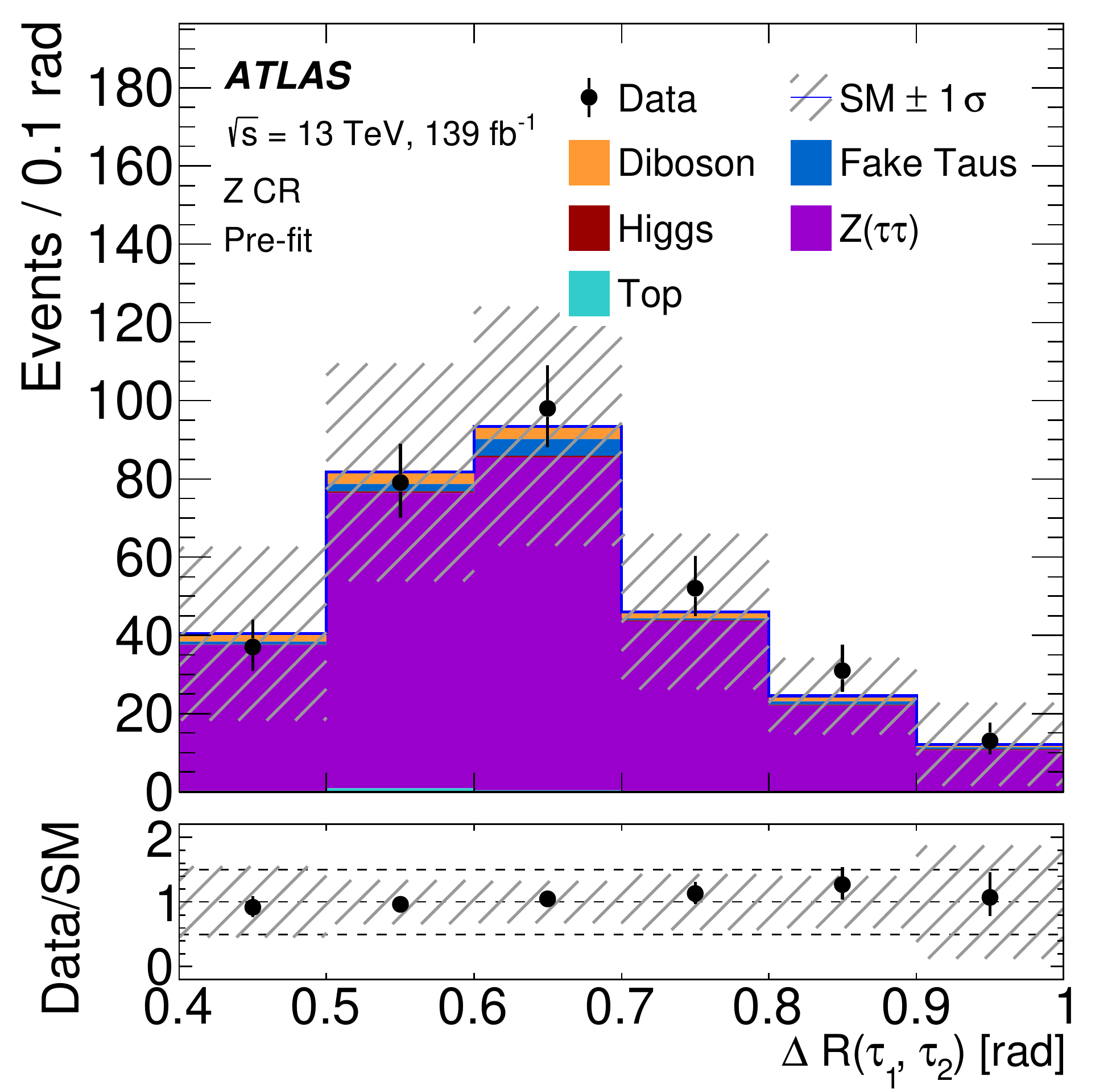}\label{fig:zjetsCR_ditau_DR}}
\hfil
\subfloat[$ m_\text{T}^{\tau_1}$ in $Z$ CR]{\includegraphics[width= 0.45\textwidth]{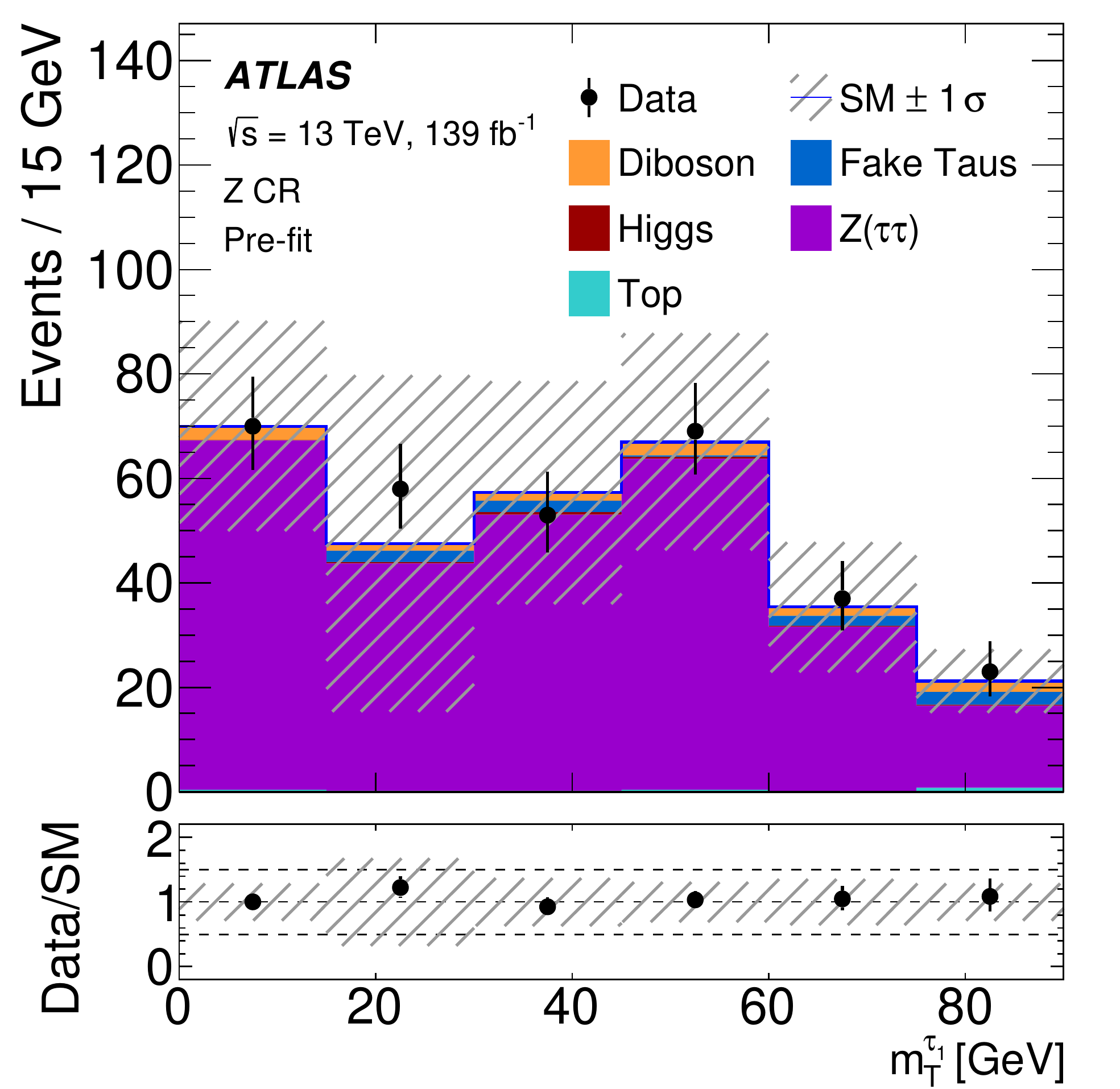}\label{fig:zjetsCR_tau_1_mtMet}}
\caption{
Kinematic distributions of some of the variables used to construct SRs are shown in the Top and $Z$ CRs:
\protect\subref{fig:topCR_ditau_mass} $m_\text{vis}(\tau_1,\tau_2)$ in Top CR,
\protect\subref{fig:topCR_tau_2_mtMet} $ m_\text{T}^{\tau_2}$ in Top CR,
\protect\subref{fig:zjetsCR_ditau_DR}    $\Delta R(\tau_1, \tau_2)$ in $Z$ CR,
\protect\subref{fig:zjetsCR_tau_1_mtMet}      $ m_\text{T}^{\tau_1}$ in $Z$ CR.
\BandIsStatSyst
The 'Other' contribution includes all the background processes not explicitly listed in the legend (\Vjets except $Z(\tau\tau)$).
\RightBinOverflow
\BottomPanelDataSM
}
\label{fig:CR_plots}
\end{figure}

Figure~\ref{fig:CR_plots} shows examples of kinematic distributions from the Top and $Z$ CRs.
The predicted background yields are shown without performing the fit and applying the resulting normalisation factors to the SM backgrounds (as described earlier in Section~\ref{sec:selection}).
Good agreement between the predicted and observed yields is seen in all plots.

\begin{table}[t]
\caption{Definitions of control and validation regions used by the analysis. The requirements are applied on top of the preselection described in Table~\ref{tab:plateau_presel}. Entries in boldface indicate selections ensuring orthogonality to other regions. A dash is used when no requirements are placed on a variable. OS and SS stand for opposite-sign and same-sign $\tau$-leptons respectively. \enquote{Comb VR} refers to an inclusive validation region containing several important backgrounds.}
 
\label{tab:crvr_selection}
\centering
\adjustbox{max width=\textwidth}{
\begin{tabular}{l|cc|c|cc|c}
\toprule
 
\textbf{Variable} & $Z$ CR & $Z$ VR  & Fake Tau VR & Top CR & Comb VR & Diboson VR \\
\midrule
Charge($\tau_1, \tau_2$) & \textcolor{black}{OS} & \textcolor{black}{OS} & \textbf{SS} &  \textcolor{black}{OS} &  \textcolor{black}{OS} & \textcolor{black}{OS} \\
 
$N_{b\textrm{-jet}}$  & \textcolor{black} 0 & \textcolor{black} 0 & \textcolor{black} 0 & \textbf{1} & \textcolor{black} 0 & \textcolor{black} 0\\
 
$\Delta R(\tau_1, \tau_2)$ &  $\color{black} <2$\phantom{$<~$} & $\color{black} <2$\phantom{$<~$} & \textbf{$>2$}\phantom{$<~$} & $>1$\phantom{$<~$} & \textbf{$>2$}\phantom{$<~$} & $\color{black} <2$\phantom{$<~$} \\
 
$m_\text{vis}(\tau_1, \tau_2)$ &  \textbf{$< 40$ \GeV} & \textcolor{black}{[40, 75] \GeV} & - & - &  $> 125$ \GeV & \textbf{[40, 75] \GeV} \\
 
$m_\text{T}^\text{tot}$ & $ >50$  \GeV & \phantom{0}$ >50$  \GeV & - & - & - & \textbf{[50, 400] \GeV}\phantom{0}\\
 
$m_\text{T}^{\tau_1}$  & - & - & - & - & - & \phantom{0}$>60$ \GeV\\
 
$m_\text{T}^{\tau_2}$  & - & - & - & - & - & \phantom{0}$>20$ \GeV \\
 
$m_\text{T}^{\tau_1} + m_\text{T}^{\tau_2}$ & - & \textbf{ $< 100$ \GeV}  & - & - & - &  $>140~\GeV$ \\
 
$\pt^{\tau_1 + \tau_2}$  & - & - & - & -  & $<125$ \GeV & - \\
 
\bottomrule
 
\end{tabular}
}
\end{table}

In the analysis phase space, approximately $80\%$ of background events with a top quark arise from \ttbar\ processes; the remaining events come from single-top processes.
Events in the Top CR are selected using the di-$\tau+\met$ triggers and are required to pass the common analysis preselection summarised in Table~\ref{tab:plateau_presel}. To define a region enriched in \ttbar\ while suppressing other common background processes such as \Zjets, the $b$-jet veto common to all other regions used in the analysis is relaxed and the events are required to have exactly one $b$-jet instead. The angular distance between the two $\tau$-leptons is required to be $\Delta R(\tau_1, \tau_2) > 1$ to suppress $Z$ boson background and to enhance the Top background purity of this CR. The two $\tau$-leptons are also required to have opposite charges to suppress contributions from fake $\tau$-leptons. The resulting Top CR has a top-quark background purity of over $90\%$, based on information from MC simulation, with over $70\%$ of events containing two true $\tau$-leptons. From simulation studies the majority of Top background events in the SRs (which all have a $b$-jet veto) are expected to have one jet that originated from a $b$-hadron but is not identified as such by the $b$-tagging algorithm. Similarly, the Top events in the Top CR are expected to have one or more jets originating from a $b$-hadron. Thus, the extrapolation from the Top CR to the SRs is physically meaningful because the same underlying processes are being considered.
 
Events from $Z$ boson production constitute the largest background in the analysis phase space, and a high-purity region is straightforward to define. Events in the $Z$ CR were recorded using the same trigger as in the SRs. The events are required to pass the common analysis preselection described in Table~\ref{tab:plateau_presel}, and to have no $b$-jets. The two $\tau$-leptons are required to have opposite charges and to satisfy $\Delta R (\tau_1, \tau_2) < 2$ and $m_\text{vis}(\tau_1, \tau_2) < 40$~\GeV\ to improve the $Z$ background purity. The events are also required to pass the $m_\text{T}^\text{tot} > 50$~\GeV\ requirement. This results in the $Z$ CR having $90\%$ purity in $Z$ boson background events.

While the SRs require $m_\text{vis}(\tau_1,\tau_2) > 40$~\GeV, the $Z$ CR demands $m_\text{vis}(\tau_1,\tau_2) < 40$~\GeV.
To validate the extrapolation to higher values of the di-$\tau$ visible invariant mass a $Z$ VR is defined. The
selection is similar to that of the $Z$ CR, but the visible invariant mass
must satisfy $40~\GeV  < m_\text{vis}(\tau_1,\tau_2) < 75~\GeV$.
To ensure orthogonality to the SRs, $m_\text{T}^{\tau_1} +
m_\text{T}^{\tau_2}$ is required to be less than $100$~\GeV.
The simulated events in the $Z$ VR are affected by large theoretical
systematic uncertainties. This is expected to be the case at large
$Z$ \pt, which is the regime where this analysis operates. Systematic
uncertainties related to the estimation of $Z$ events are discussed in Section~\ref{sec:systematics}.

The $\Delta R (\tau_1, \tau_2) > 2$ part of the analysis phase space
has reasonably large expected Top and fake $\tau$-lepton event
yields. These two background processes are closely intertwined, considering that some of the fake $\tau$-leptons originate from top-quark decays, and separating them is not practical.
Instead, an inclusive validation region Comb VR is defined, containing Top, Fake Tau, $Z$ boson, and diboson background processes.
The Comb VR is defined by requiring that events are selected by the same trigger as used in the SRs, pass the common analysis preselection and have  $\Delta R (\tau_1, \tau_2) > 2$. The charges of the two $\tau$-leptons are required to be opposite and a $b$-jet veto is applied. The Comb VR also requires $m_\text{vis}(\tau_1, \tau_2) > 125~\GeV$ and $\pt^{\tau_1 + \tau_2} < 125~\GeV$ in order to suppress potential signal contamination.
 
Diboson production is an important background process in the
SRs. However, the overall event yields from diboson processes in the
analysis phase space are too low to allow the construction of a
statistically significant CR that wouldn't also be sensitive to potential signal contributions. Instead, a validation region is defined in order to validate the normalisation of the diboson backgrounds according to their cross-sections. The events in the Diboson VR are required to pass the common SR preselection described in the upper part of Table~\ref{tab:signal_selection}. The events are also required to have $m_\text{T}^{\tau_1} > 60$~\GeV, $m_\text{T}^{\tau_2} > 20$~\GeV, and $m_\text{T}^{\tau_1} + m_\text{T}^{\tau_2} > 140$~\GeV\ to increase the diboson purity of this VR. To keep the Diboson VR orthogonal to the SRs, $m_\text{vis}(\tau_1, \tau_2) < 75~\GeV$ and $m_\text{T}^\text{tot} < 400~\GeV$ are required.
 
\begin{figure}[t]
\centering
\subfloat[$m_\text{vis}(\tau_1,\tau_2)$ in Fake Tau VR]{\includegraphics[width= 0.38\textwidth]{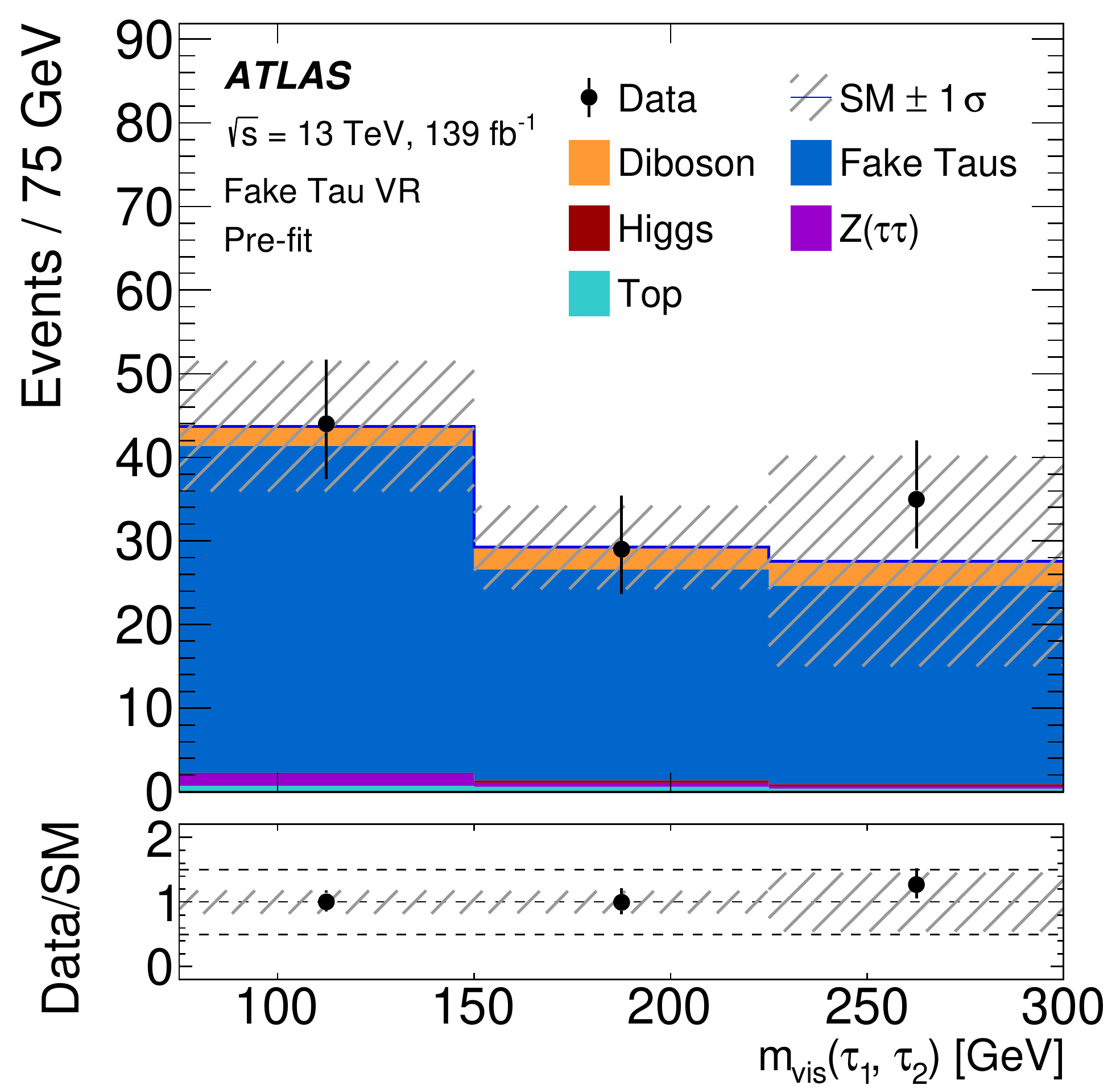}\label{fig:wjetsCR_ditau_mass}}
\hfil
\subfloat[$ m_\text{T}^{\tau_2}$ in Fake Tau VR]{\includegraphics[width= 0.38\textwidth]{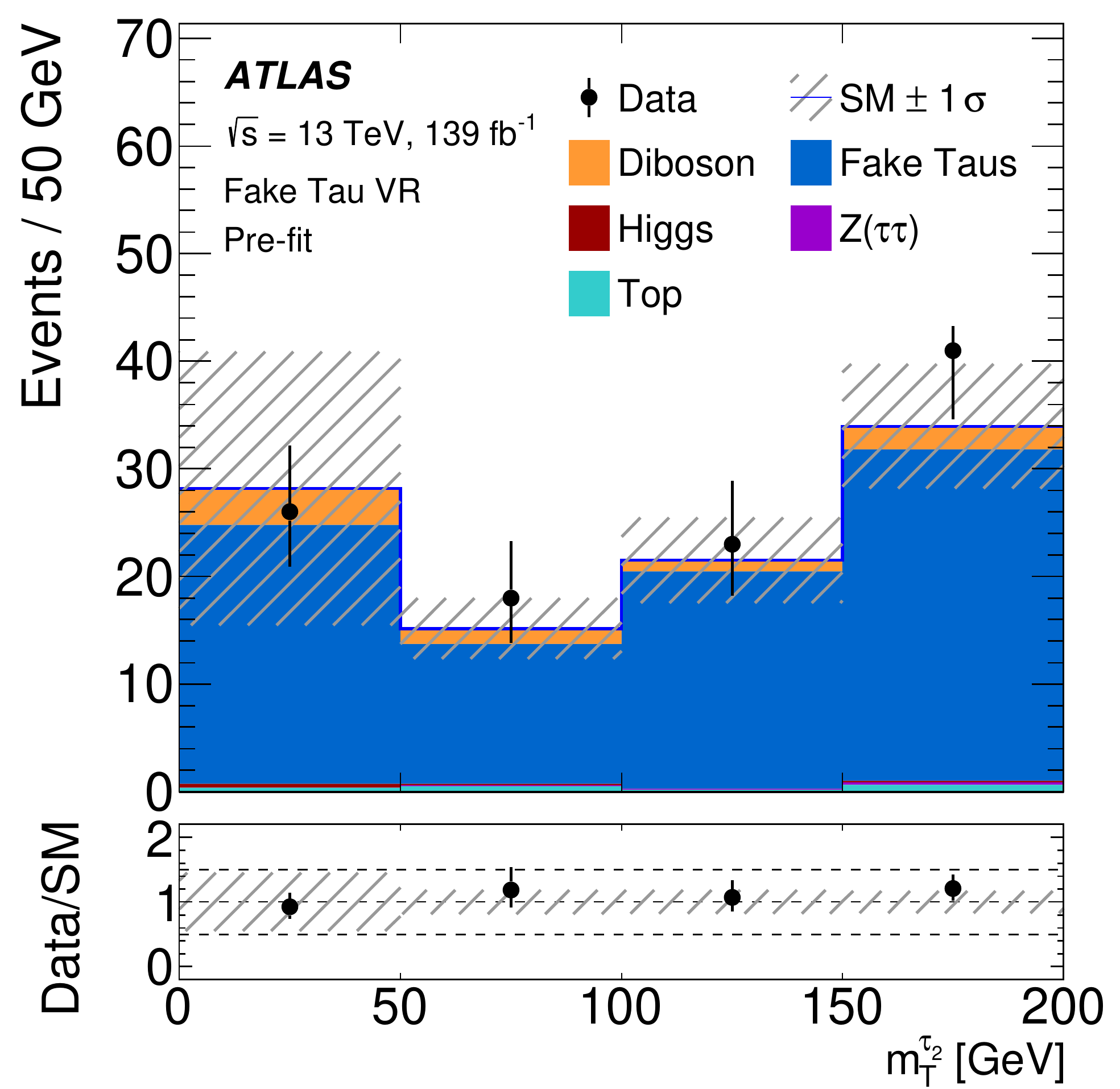}\label{fig:wjetsCR_tau_2_mtMet}}\\
\subfloat[\met in $Z$ VR]{\includegraphics[width= 0.38\textwidth]{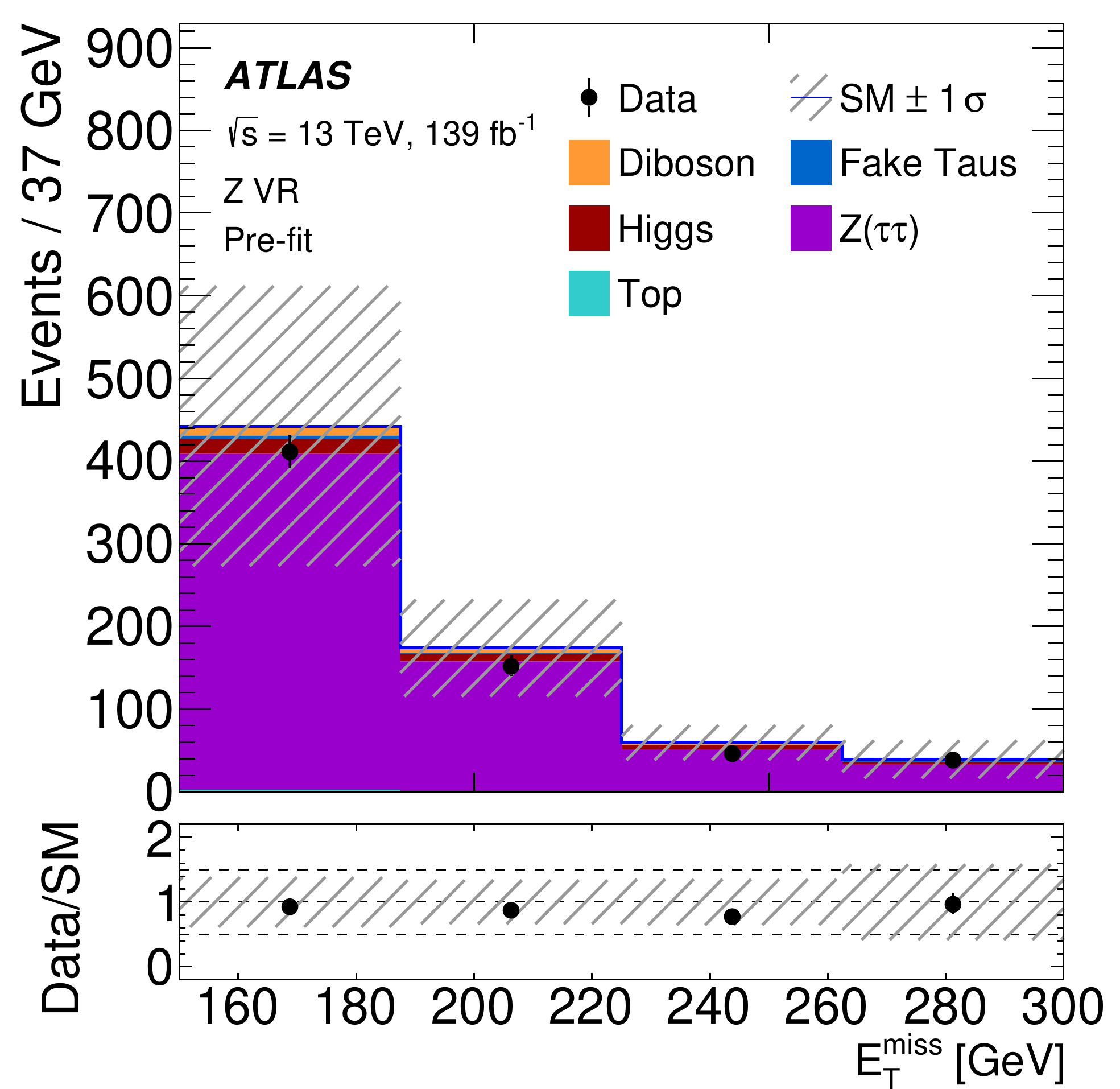}\label{fig:zjetsVR_met}}
\hfil
\subfloat[$m_\text{T}^\text{tot}$ in Comb VR]{\includegraphics[width= 0.38\textwidth]{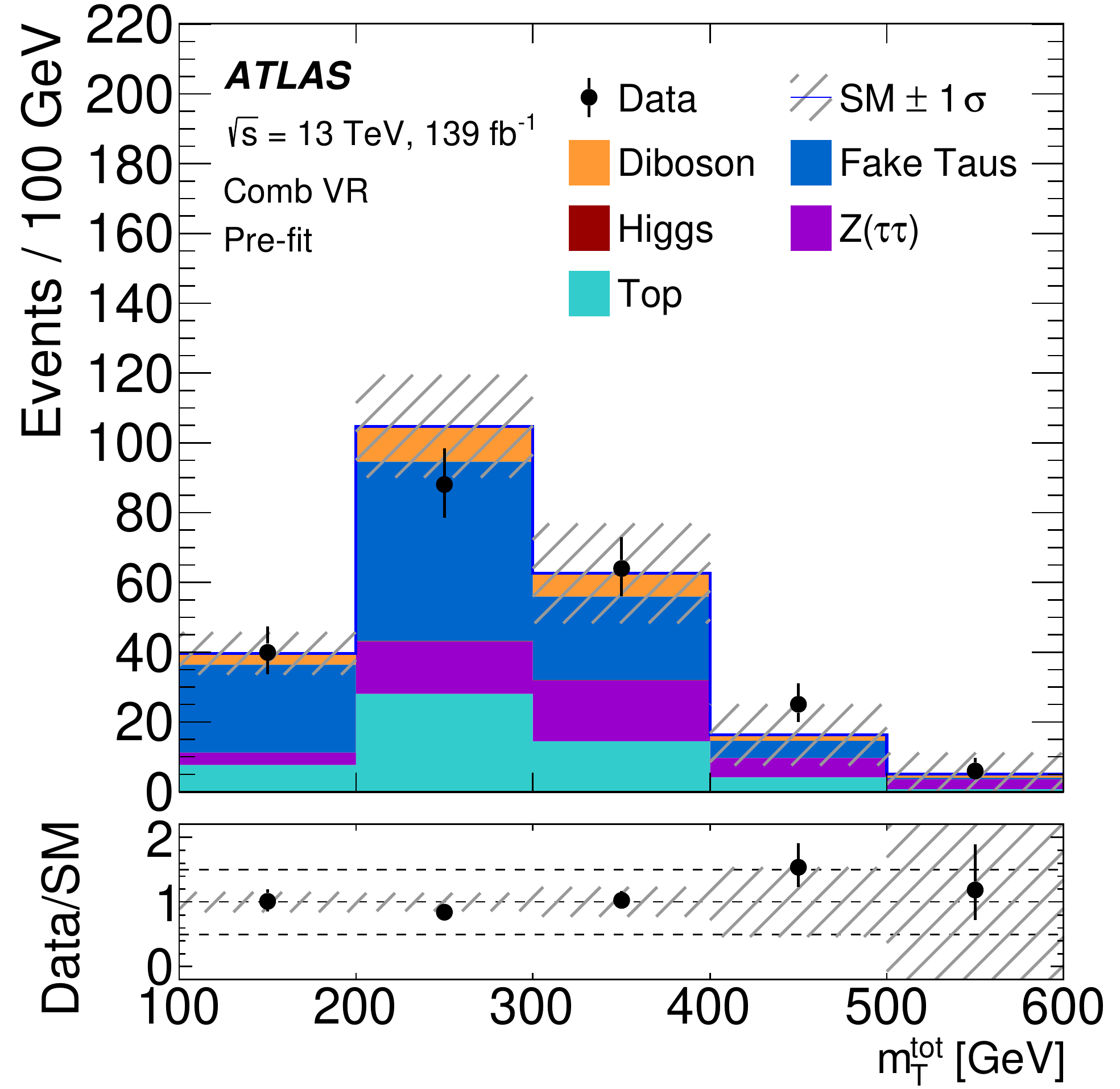}\label{fig:combVR_mTtot}}\\
\subfloat[$ m_\text{T}^{\tau_1}$ in Comb VR]{\includegraphics[width= 0.38\textwidth]{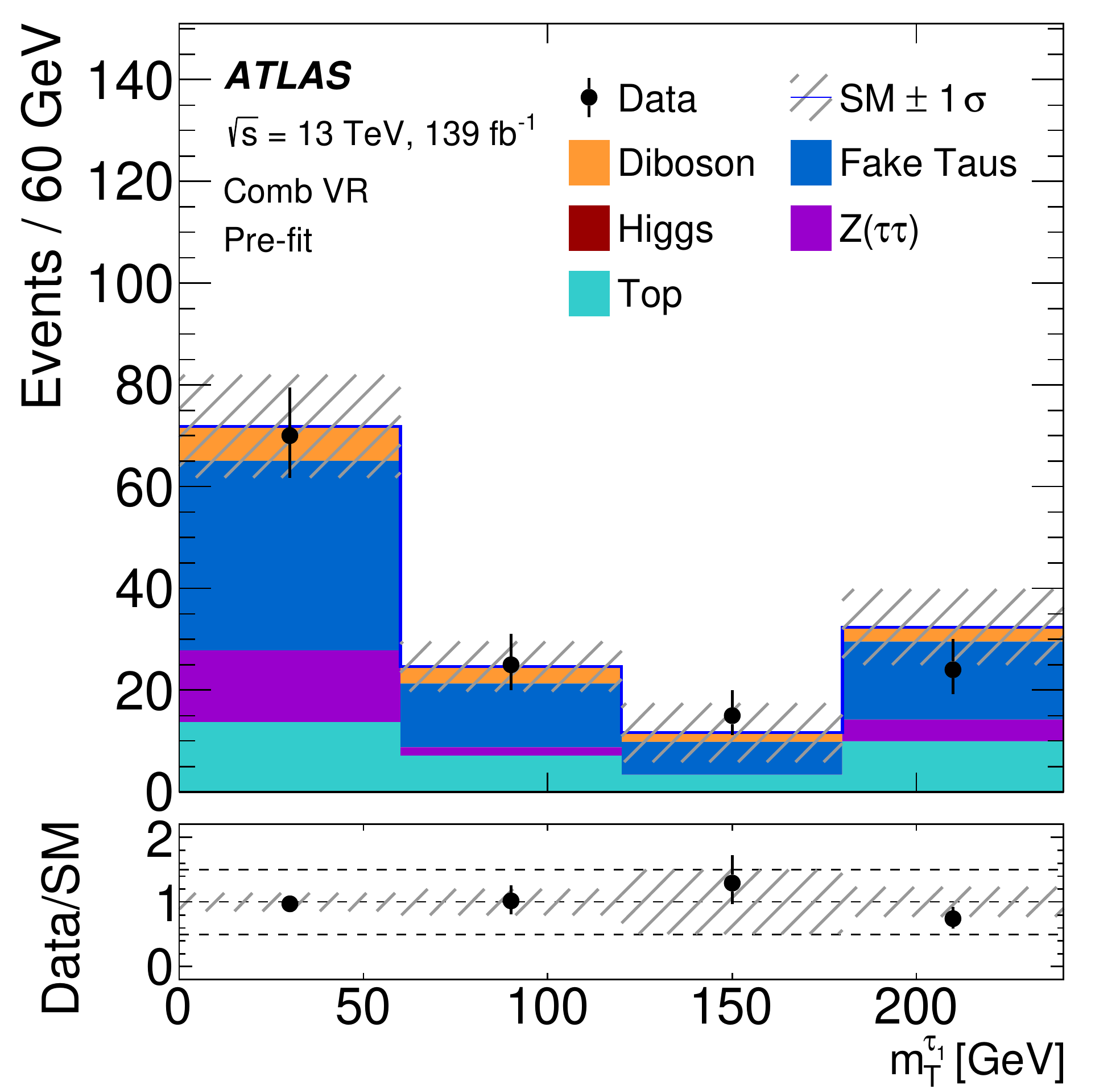}\label{fig:combVR_tau_1_mtMet}}
\hfil
\subfloat[$ m_\text{T}^{\tau_1} + m_\text{T}^{\tau_2}$ in Diboson VR]{\includegraphics[width= 0.38\textwidth]{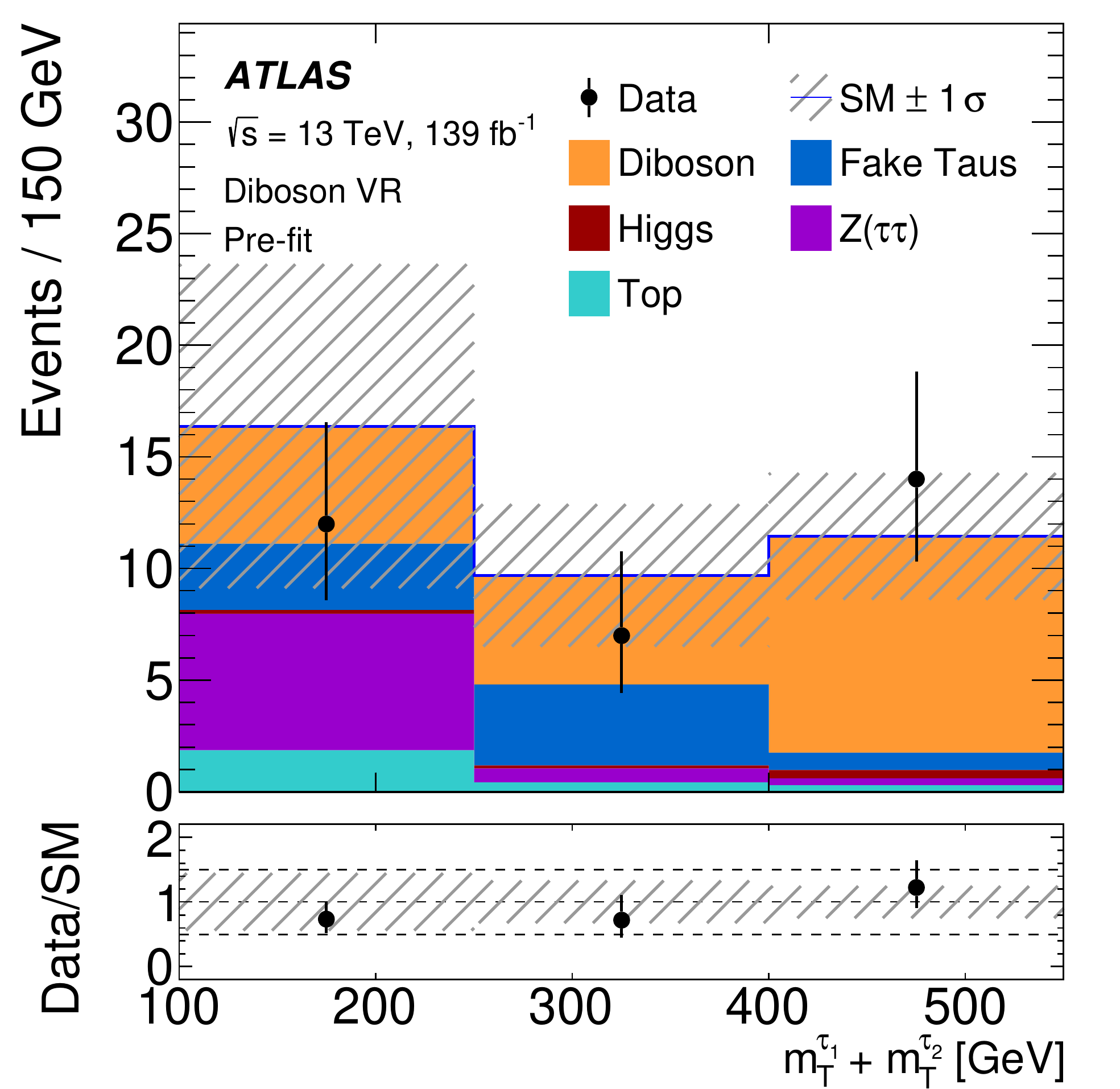}\label{fig:dibosonVR_tau_1_mtMet_tau_2_mtMet}}
\caption{ Kinematic distributions of some of the variables used to construct SRs are shown in the four VRs:
\protect\subref{fig:wjetsCR_ditau_mass} $m_\text{vis}(\tau_1,\tau_2)$ in Fake Tau VR,
\protect\subref{fig:wjetsCR_tau_2_mtMet} $ m_\text{T}^{\tau_2}$ in Fake Tau VR,
\protect\subref{fig:zjetsVR_met} \met in $Z$ VR,
\protect\subref{fig:combVR_mTtot} $m_\text{T}^\text{tot}$ in Comb VR,
\protect\subref{fig:combVR_tau_1_mtMet}      $ m_\text{T}^{\tau_1}$ in Comb VR,
\protect\subref{fig:dibosonVR_tau_1_mtMet_tau_2_mtMet} $ m_\text{T}^{\tau_1} + m_\text{T}^{\tau_2}$ in Diboson VR.
\BandIsStatSyst
\RightBinOverflow
\BottomPanelDataSM
}
\label{fig:VR_plots}
\end{figure}

Figure~\ref{fig:VR_plots} shows examples of kinematic distributions from the four VRs used by the analysis.
The predicted background yields are shown without the derivation and extrapolation of normalisation factors in the CRs (as described in Section~\ref{sec:selection}).
The maximum contribution from potential signal models with the parameter choices considered by the analysis does not exceed $1\%$ in the CRs and $10\%$ in the VRs. The only exception is the Diboson VR, where the potential signal contribution can reach $35\%$ of the total yield. 
 
\FloatBarrier
 
\section{Systematic uncertainties}
\label{sec:systematics}

The analysis considers theoretical, experimental and statistical
uncertainties in signal and SM background modelling. The
uncertainties are included in the likelihood fits as nuisance
parameters with Gaussian probability densities. Theoretical
uncertainties comprise generator-modelling-related uncertainties,
cross-section uncertainties, and uncertainties related to the choice
of PDF set. Such uncertainties are assumed to be uncorrelated
between background processes.
Systematic uncertainties in the reconstruction, identification,
corrections and calibrations of various analysis objects are referred to as experimental uncertanties. These uncertainties are assumed to be correlated across processes and analysis regions.

Jet-related experimental uncertainties include uncertainties in the energy scale \cite{PERF-2016-04} and resolution~\cite{JETM-2018-05}, as well as jet-vertex-tagging uncertainties \cite{ATLAS-CONF-2014-018} and flavour-tagging uncertainties \cite{FTAG-2018-01,ATLAS-CONF-2018-001,ATLAS-CONF-2018-006}. The experimental uncertainties related to $\tau$-leptons include uncertainties from the reconstruction and identification algorithms as well as energy scale and electron veto uncertainties \cite{ATL-PHYS-PUB-2015-045,ATLAS-CONF-2017-029}. Uncertainties related to electrons \cite{EGAM-2018-01} and muons \cite{PERF-2015-10} are negligible. 
Uncertainties in the \met are estimated by taking into account the uncertainties related to the energy or momentum of calibrated objects as well as the uncertainties affecting the soft-term contributions \cite{ATLAS-CONF-2018-023}.
 
The uncertainty in the combined \RunTwo integrated luminosity is $1.7\%$ \cite{ATLAS-CONF-2019-021}, obtained from luminosity measurements with the LUCID-2
detector \cite{Avoni:2018}. The \pileup reweighting applied to
simulated events to match the observed conditions in data is assigned systematic uncertainties, reaching at most $6.4\%$. Uncertainties related to the trigger efficiency correction factors are at most $4\%$ depending on the \met\ of the event. All of the mentioned systematic uncertainties are applied to all simulated samples.
 
Systematic uncertainties related to the fake-$\tau$ modelling consist of uncertainties in the number of events in the regions used to measure the fake factors and in the regions where they are applied, the statistical and systematic uncertainties of the MC samples in the regions where fake factors are applied, and the uncertainties from the extrapolation and application of fake factors. Based on a comparison of different correlation schemes the correlation effects are found to be negligible and these uncertainties are considered to be uncorrelated across regions.

Theoretical uncertainties related to variations of the PDFs are computed for all background and signal samples.
The effect of using the nominal PDF and $100$ variations is represented by a set of generator weights. The standard deviation of all the variations is computed and used as the PDF uncertainty. The uncertainties related to the value of the strong coupling constant \alphas\ are estimated by halving the difference of two PDF sets evaluated with $\alphas = 0.117$ and $\alphas = 0.119$. The PDF and \alphas\ systematic uncertainties are then combined in quadrature, following the \PDFforLHC recommendations~\cite{Butterworth:2015oua}. To estimate the uncertainties due to missing higher-order corrections, the values of renormalisation and factorisation scales \muR and \muF~\cite{ATL-PHYS-PUB-2018-009} are varied independently by a factor of $2$, requiring $0.5 \leq \muR/\muF \leq 2$. The envelope of the effects of these scale variations is used to estimate the effect of the scale uncertainty. The cross-section uncertainty is applied to all background processes that are not normalised to data in the likelihood fits.
 
For the simulated $Z$ and $W$ boson samples, additional uncertainties are estimated by varying the resummation and CKKW matching scales~\cite{Catani:2001cc,Hoeche:2009rj}. For the \Ztautaujets and diboson backgrounds, comparisons of the nominal \Sherpa-produced samples with \MGNLOPY-produced samples are used to estimate the uncertainties.
For the \ttbar\ and single-top-quark production processes, comparisons of the nominal \POWPY samples with \MGNLOPY samples are used to estimate uncertainties related to hard scattering. Similarly, parton showering uncertainties are evaluated by comparing the \POWPY samples with \POWHER[7] samples. Uncertainties related to initial-state and final-state radiation are also considered for \ttbar and single-top-quark production. Finally, for single-top-quark production samples, the effect of the diagram removal scheme \cite{Frixione:2008yi} is compared with the effect of the diagram subtraction scheme~\cite{Frixione:2008yi,ATL-PHYS-PUB-2016-020} to parameterise the uncertainty in the treatment of the $Wt/\ttbar$ interference.

For the \TwoHDMa model signal samples the theoretical uncertainties related to the PDFs and renormalisation and factorisation scales are considered. Additional experimental uncertainties account for the use of the simplified parameterised calorimeter response instead of the full $\GEANT$-based detector simulation.

\begin{table}[t]
\caption{
Systematic uncertainties in the post-fit SM background prediction in the six SR bins.
The normalisation uncertainty includes the effect of keeping the normalisation factors as free-floating parameters in the fit.
\enquote{Other} includes the uncertainties
arising from trigger efficiency corrections, electrons, muons, modelling of \pileup,
and the \met computation.
The individual uncertainties can be correlated and do not
necessarily add in quadrature to equal the total uncertainty.
}
\label{tab:SRsyst}
\centering
\begin{tabular}{lrrrrrr}
\toprule
Region & Low$_{m_A}$ & Low$_{m_A}$ & Low$_{m_A}$ & Low$_{m_A}$ & High$_{m_A}$ & High$_{m_A}$ \\
$m_\text{T}^{\tau_1} + m_\text{T}^{\tau_2}$ [$\GeV$]       &     $\in [100,250]$  & $\in [250,400]$ & $\in [400,550]$  & $>550$  & $\in [400,750]$  & $> 750$   \\
\midrule
Theoretical                     & $15.9\%$& $20.9\%$& $15.4\%$& $13.3\%$ & $14.2\%$& $21.7\%$ \\
Fake $\tau$-leptons             & $6.2\%$ & $19.1\%$& $6.0\%$ & $3.0\%$  & $4.0\%$ & $13.2\%$ \\
Jets                            & $6.2\%$ & $7.9\%$ & $5.2\%$ & $11.0\%$ & $3.4\%$ & $7.9\%$ \\
True $\tau$-leptons             & $1.6\%$ & $3.1\%$ & $7.0\%$ & $10.6\%$ & $4.8\%$ & $5.0\%$ \\
Normalisation                & $4.5\%$ & $4.0\%$ & $4.8\%$ & $8.1\%$  & $4.8\%$ & $6.1\%$ \\
MC statistical                  & $7.6\%$ & $13.2\%$& $9.3\%$ & $15.6\%$ & $9.2\%$ & $22.2\%$ \\
Cross-section                   & $2.7\%$ & $4.7\%$ & $10.0\%$& $9.8\%$  & $11.3\%$& $8.5\%$ \\
Other                           & $4.1\%$ & $2.7\%$ & $5.3\%$ & $6.3\%$  & $4.9\%$ & $4.5\%$ \\
\midrule
Total                           & $20.8\% $  & $ 32.8\% $ & $23.8\%$ & $ 28.3\% $ & $22.2\%$ & $36.3\%$ \\
\bottomrule
\end{tabular}
 
\end{table}
 
A breakdown of the different sources of uncertainty in the SRs is presented in Table~\ref{tab:SRsyst}. The largest contribution comes from a combination of various theoretical uncertainties -- comparisons with alternative generators for diboson, $Z$ boson, and \ttbar\ production processes, as well as PDF and scale variations. Other significant sources of uncertainty are the limited number of MC events and uncertainties in the modelling of fake $\tau$-leptons.


\FloatBarrier
 
\section{Results}
\label{sec:result}

\begin{figure}[t]
\centering
\includegraphics[width=0.8\textwidth]{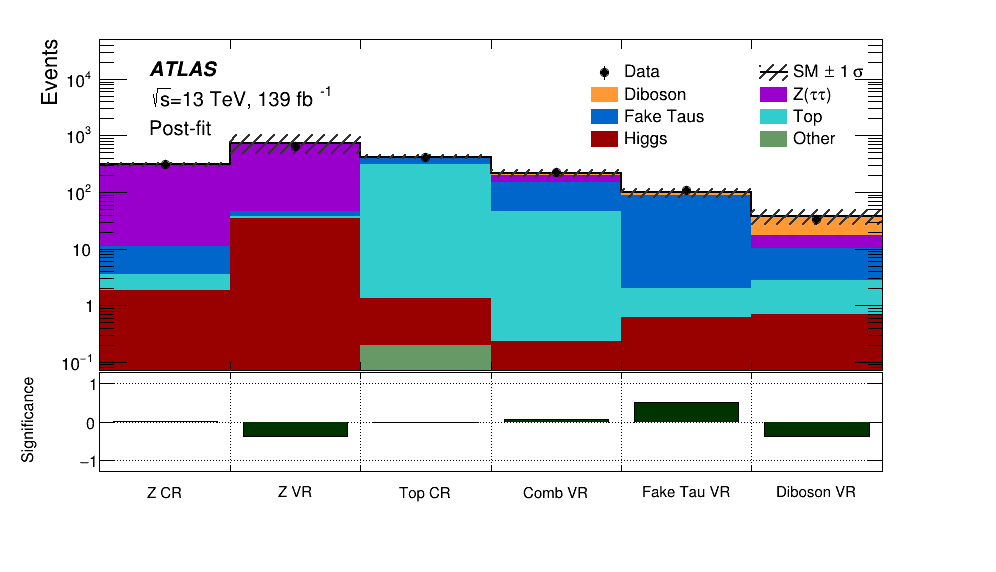}
\caption{
A comparison of the observed and expected yields in each of the two control regions and four validation regions used by the analysis is shown in the upper panel. The lower panel shows the statistical significance of the observation given the number of predicted events. The 'Other' contribution includes all the background processes not explicitly listed in the legend (\Vjets except $Z(\tau\tau)$).
\Postfit
\BandIsStatSyst
}
\label{fig:CRVR_plots}
\end{figure}

 
Figure~\ref{fig:CRVR_plots} summarises the observed and expected yields in all CRs and VRs used by the analysis.
The normalisation factors determined by the background-only fit to the data in the CRs and propagated to the VRs and SRs to apply corrections to the SM backgrounds are $\omega_\text{top} = 0.82 \pm 0.15$ and $\omega_\text{Z} = 1.04 \pm 0.22$.
All four validation regions show agreement within one standard deviation between the SM prediction and the data, signifying good modelling of the SM backgrounds. 
 
\begin{figure}[t]
\centering
\includegraphics[width=0.8\textwidth]{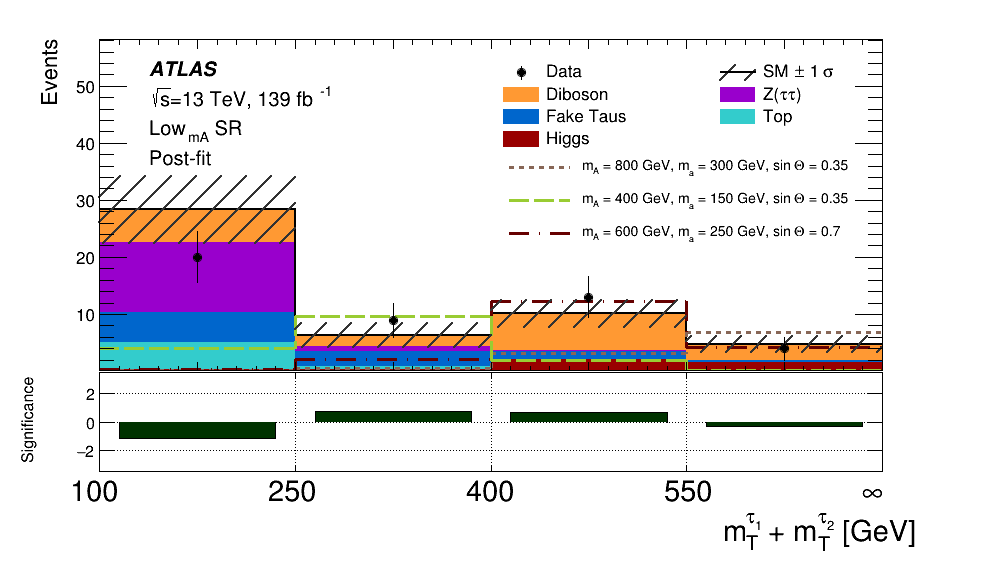}
\caption{A comparison of the observed and expected yields in the four bins of the Low$_{m_A}$ SR region used in the analysis is shown in the upper panel. The lower panel shows the statistical significance of the observation given the predicted number of events.
\Postfit
\BandIsStatSyst
}
\label{fig:lowmasr_pull}
\end{figure}

\begin{figure}[t]
\centering
\includegraphics[width=0.8\textwidth]{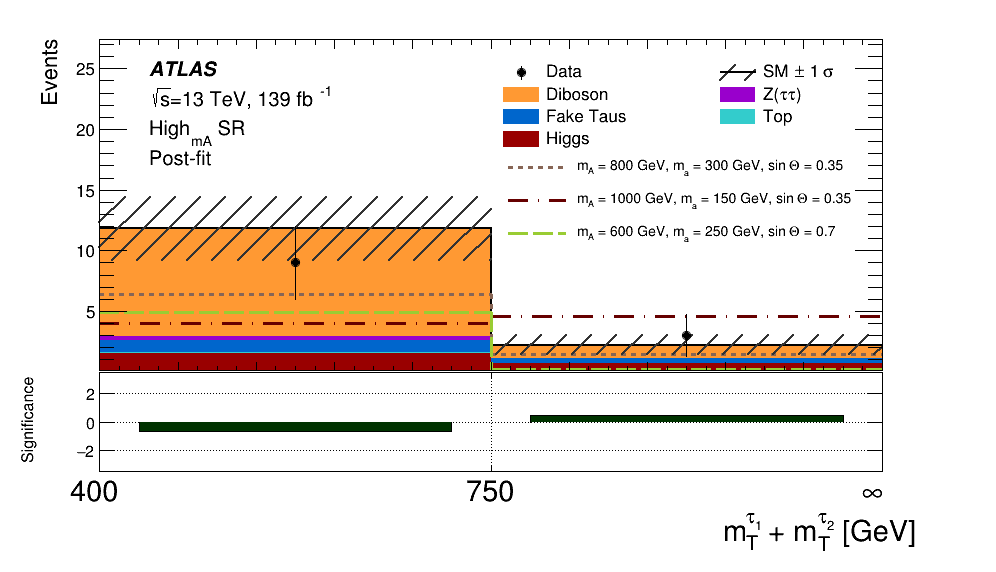}
\caption{A comparison of the observed and expected yields in the two bins of the High$_{m_A}$ SR region used in the analysis is shown in the upper panel. The lower panel shows the statistical significance of the observation given the predicted number of events.
\Postfit
\BandIsStatSyst
}
\label{fig:highmasr_pull}
\end{figure}

The yields in each bin of the Low$_{m_A}$ SR are summarised in Table~\ref{tab:srlow_yield}.
Comparisons of observed and predicted yields and the corresponding statistical significances \cite{ATL-PHYS-PUB-2020-025} are shown in Figure~\ref{fig:lowmasr_pull} together with the benchmark parameter configurations of the signal model predictions. These signal benchmarks represent different kinematic regimes that the Low$_{m_A}$ SR is sensitive to.
The yields in the High$_{m_A}$ SR bins are summarised in Table~\ref{tab:srhigh_yield} and Figure~\ref{fig:highmasr_pull} together with the corresponding statistical significances and benchmark parameter configurations of the signal model. No significant excess of data over the predicted SM backgrounds is observed in any of the bins.

\begin{table}[t]
\caption{Observed and expected yields in the Low$_{m_A}$ SR
bins after the background-only fit. The upper part of the table describes the breakdown of
predicted SM backgrounds. The lower part of the table shows predicted yields of three signal model benchmark parameter configurations.}
\label{tab:srlow_yield}
\centering
\begin{tabular}{lcccc}
\toprule
$m_\text{T}^{\tau_1} + m_\text{T}^{\tau_2}$       &     $\in [100,250]$ \GeV       & $\in [250,400]$ \GeV            & $\in [400,550]$ \GeV            & $>550$ \GeV              \\
\midrule
Observed events          & $20$              & $9$              & $13$              & $4$                    \\
\midrule
Fitted background events     & $28.3 \pm 5.9$\phantom{0}          & $6.3 \pm 2.1$          & $10.1 \pm 2.4$\phantom{0}          & $4.8 \pm 1.3$              \\
\midrule
Diboson   & $5.7 \pm 2.4$          & $2.0 \pm 0.7$          & $6.5 \pm 2.1$          & $2.8 \pm 1.1$              \\
$Z(\tau \tau)$  & $12.3 \pm 3.8$\phantom{0}          & $0.8_{-0.8}^{+1.0}$\,          & $0.1_{-0.1}^{+0.2}$\,          & $0.1_{-0.1}^{+0.2}$\,              \\
$t\bar{t}$    & $4.4 \pm 2.7$          & $0.3_{-0.3}^{+0.8}$\,          & $0.0 \pm 0.0$          & $0.0 \pm 0.0$              \\
Single-top    & $0.2_{-0.2}^{+0.4}$\,          & $0.2_{-0.2}^{+0.3}$\,          & $0.0 \pm 0.0$          & $0.0 \pm 0.0$              \\
$t\bar{t} V$   & $0.1 \pm 0.0$          & $0.1 \pm 0.1$          & $0.1 \pm 0.0$          & $0.1 \pm 0.0$              \\
Higgs        & $0.4 \pm 0.1$          & $0.4 \pm 0.1$          & $2.0 \pm 0.4$          & $1.5 \pm 0.4$              \\
Fake tau     & $5.3 \pm 1.8$          & $2.6 \pm 1.3$          & $1.4 \pm 0.6$          & $0.3 \pm 0.2$              \\
\midrule
$m_a = 150$ \GeV, $\sin\Theta = 0.35$ & \multirow{2}{*}{$4.1 \pm 0.6$} & \multirow{2}{*}{$9.6 \pm 1.2$} & \multirow{2}{*}{$1.9  \pm 0.4$} & \multirow{2}{*}{$0.1  \pm 0.0$} \\
$m_A = 400$ \GeV, $\tan\beta = 1$ & & & & \\[5pt]
$m_a = 250$ \GeV, $\sin\Theta = 0.7$  & \multirow{2}{*}{$0.4 \pm 0.1$} & \multirow{2}{*}{$2.0 \pm 0.3$} & \multirow{2}{*}{$12.3 \pm 1.3$} & \multirow{2}{*}{$4.3  \pm 0.6$}\\
$m_A = 600$ \GeV, $\tan\beta = 1$ & & & & \\[5pt]
$m_a = 300$ \GeV, $\sin\Theta = 0.35$ &\multirow{2}{*}{$0.2 \pm 0.0$} & \multirow{2}{*}{$0.5 \pm 0.1$} & \multirow{2}{*}{$3.1 \pm 0.4$} & \multirow{2}{*}{$6.8 \pm 0.9$} \\
$m_A = 800$ \GeV, $\tan\beta = 1$ & & & & \\
 
\bottomrule
\end{tabular}
\end{table}

\begin{table}[t]
\caption{Observed and expected yields in the High$_{m_A}$ SR bins after the background-only fit. The upper part of the table describes the breakdown of predicted SM backgrounds. The lower part of the table shows predicted yields of three signal model benchmark parameter configurations.}
\label{tab:srhigh_yield}
\centering
\begin{tabular}{lcc}
\toprule
$m_\text{T}^{\tau_1} + m_\text{T}^{\tau_2}$       & $\in [400,750]$ \GeV            & $> 750$ \GeV              \\
\midrule
Observed events   & $9$              & $3$                    \\
\midrule
Fitted background events & $11.8 \pm 2.6$\phantom{0}          & $2.2 \pm 0.8$              \\
\midrule
Diboson        & $8.9 \pm 2.4$          & $1.1 \pm 0.6$              \\
$Z(\tau \tau)$   & $0.4_{-0.4}^{+0.5}$\,          & $0.0_{-0.0}^{+0.1}$\,              \\
$t\bar{t} V$  & $0.1 \pm 0.0$          & $0.1 \pm 0.0$              \\
Higgs         & $1.5 \pm 0.3$          & $0.7 \pm 0.2$              \\
Fake tau    & $0.9 \pm 0.5$          & $0.4 \pm 0.3$              \\
\midrule
$m_a = 150$ \GeV, $\sin\Theta = 0.35$ &\multirow{2}{*}{$4.0  \pm 0.5$} & \multirow{2}{*}{$4.6 \pm 0.6$} \\
$m_A = 1000$ \GeV, $\tan\beta = 1$ & & \\[5pt]
$m_a = 250$ \GeV, $\sin\Theta = 0.7$ & \multirow{2}{*}{$4.9  \pm 0.7$} & \multirow{2}{*}{$0.2 \pm 0.1$} \\
$m_A = 600$ \GeV, $\tan\beta = 1$ & & \\[5pt]
$m_a = 300$ \GeV, $\sin\Theta = 0.35$ &\multirow{2}{*}{$6.4  \pm 0.8$} & \multirow{2}{*}{$1.4 \pm 0.3$} \\
$m_A = 800$ \GeV, $\tan\beta = 1$ & & \\
\bottomrule
\end{tabular}
\end{table}
 
For every signal model parameter configuration considered by the analysis a signal-plus-background fit is performed. The fit includes the CRs, and the Low$_{m_A}$ or High$_{m_A}$ SR. The contribution from the signal model is included as a free-floating parameter. The fit is performed simultaneously over all the CRs and SR bins.
The probability that the observed yields in the Low$_{m_A}$ SR or High$_{m_A}$ SR are compatible with the signal-plus-background hypotheses are computed. A one-sided profile-likelihood-ratio test statistic as implemented in the \histfitter\ framework is used. The results are obtained using the \CLs prescription~\cite{Read:2002hq} and presented as exclusion contours at $95\%$ confidence level (CL) in the two-dimensional parameter spaces.

\begin{figure}[t]
\centering
\subfloat[]{\includegraphics[width= 0.5\textwidth]{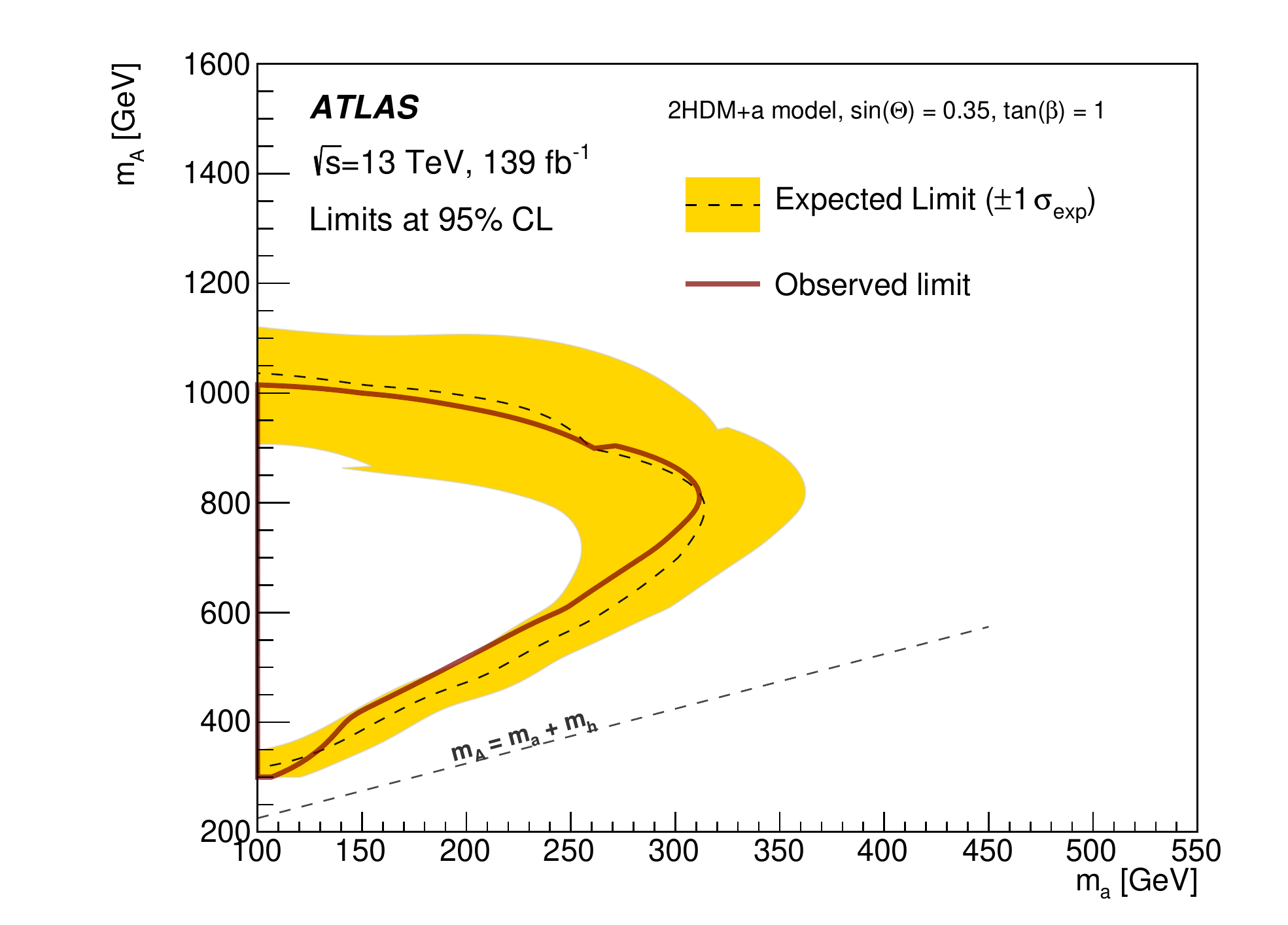}\label{fig:contour_plot}}
\hfil
\subfloat[]{\includegraphics[width= 0.5\textwidth]{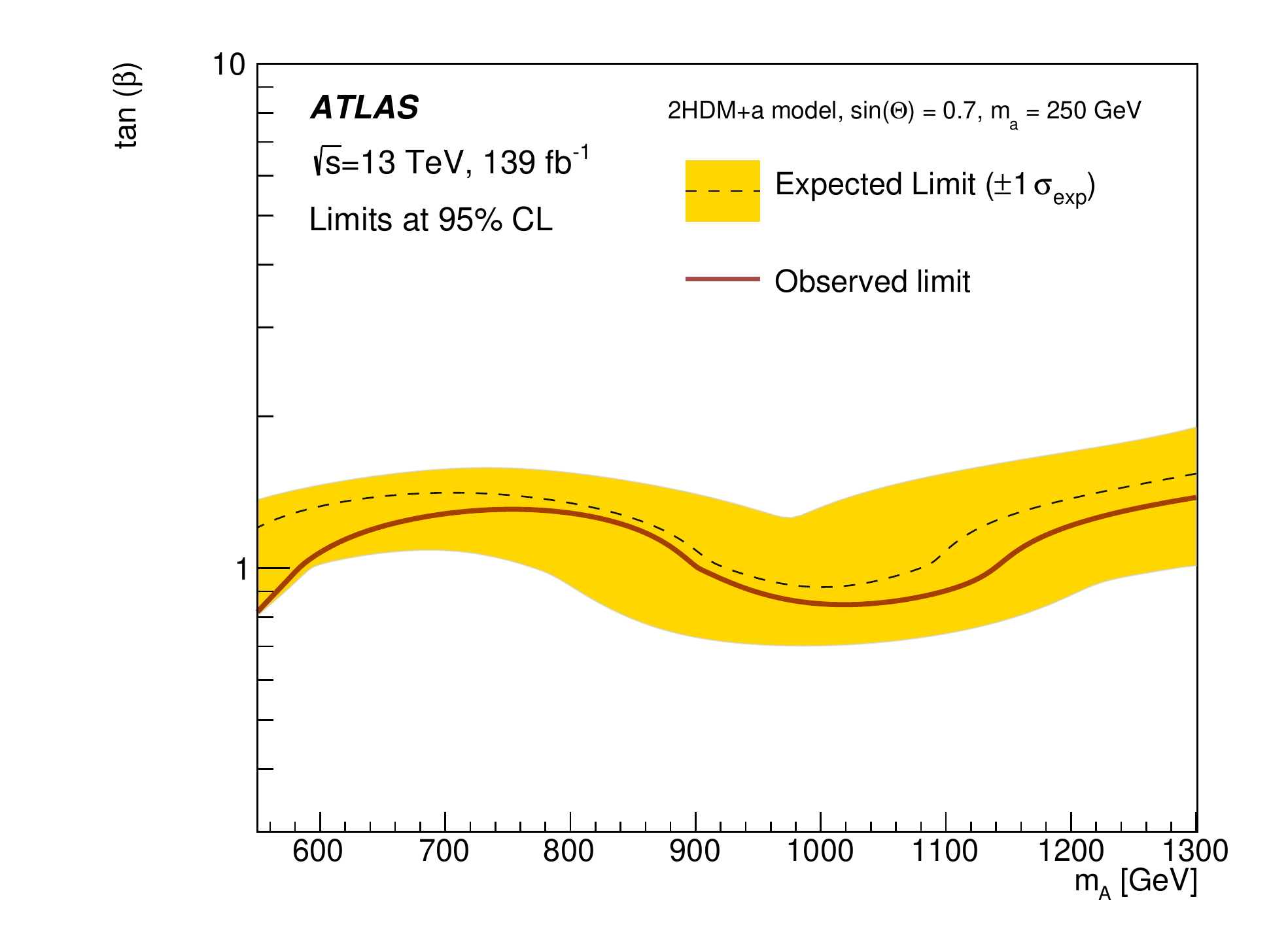}\label{fig:contour_plot2}}\\
\caption{Observed and expected exclusion contours at $95\%$ CL
\protect\subref{fig:contour_plot} as a function of $m_a$ and $m_A$ with $\tan{\beta} = 1 $ and $\sin{\Theta} = 0.35$, and
\protect\subref{fig:contour_plot2} as a function of $m_A$ and $\tan{\beta}$ with $m_a = 250$~\GeV\ and $\sin{\Theta} = 0.7$.
The excluded area in \protect\subref{fig:contour_plot} is to the left of the solid line, and in \protect\subref{fig:contour_plot2} below the solid line.
}
\end{figure}

The first parameter space is defined by setting $\tan\beta = 1$ and
$\sin\Theta = 0.35$, while varying the values of $m_a \in
[100,400]~\GeV$ and $m_A \in [300,1200]$~\GeV. The Low$_{m_A}$ SR is
sensitive to models with $m_A \leq 800~\GeV$, while the High$_{m_A}$ SR is
more sensitive to the $m_A \geq 1000$~\GeV\ part of the parameter
space. For the Low$_{m_A}$ SR, computations of \CLs rely on the
asymptotic properties of the profile-likelihood
ratio \cite{Cowan:2010js}. For the High$_{m_A}$ SR however, the asymptotic
approximation is not valid because of the low yields in one of the regions,
so Monte Carlo pseudo-experiments are used instead.
The exclusion contours are derived for the Low$_{m_A}$ SR and High$_{m_A}$ SR separately and then combined
using a method based on expected \CLs values,
as the two SRs are not statistically independent.
The observed data yields are higher than the predicted SM backgrounds in the $m_\text{T}^{\tau_1} + m_\text{T}^{\tau_2} \in [250,550]$~\GeV\ bins of the Low$_{m_A}$ SR, resulting in the observed limits being weaker than the expected limits. Pseudoscalar singlets $a$ with masses up to $300$~\GeV\ are excluded for $m_A = 800~\GeV$. The combined exclusion contour is shown in Figure~\ref{fig:contour_plot}.
 
The second two-dimensional parameter space considered by the analysis
spans $m_A \in [500,1300]$~\GeV\ and $\tan\beta \in [0.3, 20]$
with the parameters $m_a$ and $\sin\Theta$ fixed to the values of $250$ \GeV\ and $0.7$ respectively.
Only the Low$_{m_A}$ SR has significant sensitivity to the signal model with these
parameter value choices. Monte Carlo pseudo-experiments are used for the
computation of the exclusion limits because the asymptotic approximation is not appropriate for
all parameter configurations of the signal model. Since in the Low$_{m_A}$ SR the data yields are
higher than expected in the $m_\text{T}^{\tau_1} +
m_\text{T}^{\tau_2} \in [250,550]$~\GeV\ bins, the observed limits are
less stringent than the expected ones for the parameter configurations of the signal model that these bins are sensitive to. This corresponds to models with lower values of $m_A$.
Signal points in the parameter space with $\tan\beta \leq 1$ are excluded for $m_A < 900$~\GeV. The exclusion contour is shown in Figure~\ref{fig:contour_plot2}.

For each of the four bins in the Low$_{m_A}$ SR and two bins in the High$_{m_A}$ SR, model-independent upper limits are set on the cross-sections of potential BSM processes. A signal-plus-background fit over all the CRs and the selected SR bin is performed. A generic signal contribution with free-floating normalisation is assumed in the SR, but not in the CRs. The profile-likelihood-ratio test statistic is evaluated using the \histfitter\ framework. Upper limits of $0.4{-}0.8$~fb are placed on the visible cross-section \sigvis at $95\%$ CL. Upper limits on the number of signal events given the observed number of events and predicted number of background events, confidence levels for the background-only hypotheses, the discovery $p$-value, and the associated significances in all of the SR bins are summarised in Table~\ref{tab:discSR}.

\begin{table}[t]
\caption{
Upper limits at $95\%$ CL on the visible cross-section \sigvis,
on the number of signal events ($S_\mathrm{obs}^{95}$),
and on the number of signal events given the expected number (and $\pm 1\sigma$ excursions of the expectation) of background events ($S_\mathrm{exp}^{95}$).
The last two columns indicate the \CLb value (i.e. the confidence level observed for the background-only hypothesis),
the discovery $p$-value ($p_0$),
and its associated significance $Z$.
}
\label{tab:discSR}
\centering
\begin{tabular}{lccccc}
\toprule{Signal region} & \sigvis[fb]  &  $S_\mathrm{obs}^{95}$  & $S_\mathrm{exp}^{95}$ & \CLb & $p_0$ ($Z$)  \\
\midrule
Low$_{m_A}$ SR & & & & & \\
$m_\text{T}^{\tau_1} + m_\text{T}^{\tau_2}$ & & & & & \\
\midrule
$  [100,250]$ \GeV  & $0.08$ &  $10.7$\phantom{0} & $ { 12.5 }^{ +5.2 }_{ -3.5 }$\phantom{0} & $0.27$ & $ 0.86$~$(-1.07)$ \\
$ [250,400]$ \GeV & $0.07$ &  $9.1$ & $ { 7.6 }^{ +3.1 }_{ -1.6 }$ & $0.72$ & $ 0.30$~\phantom{$-$}$(0.53)$ \\
$ [400,550]$ \GeV & $0.08$ &  $10.8$\phantom{0} & $ { 8.9 }^{ +3.4 }_{ -2.3 }$ & $0.75$ & $ 0.26$~\phantom{$-$}$(0.65)$ \\
$ > 550$ \GeV     & $0.04$ &  $5.8$ & $ { 6.0 }^{ +2.6 }_{ -1.6 }$ & $0.42$ & $ 0.61$~$(-0.29)$ \\
\midrule
High$_{m_A}$ SR & & & & & \\
$m_\text{T}^{\tau_1} + m_\text{T}^{\tau_2}$ & & & & & \\
\midrule
$ [400,750]$ \GeV & $0.05$ &  $7.6$ & $ { 8.8 }^{ +3.1 }_{ -2.4 }$ & $0.34$ & $ 0.85$~$(-1.03)$ \\
$ > 750$ \GeV  & $0.04$ &  $5.4$ & $ { 4.6 }^{ +1.8 }_{ -0.8 }$ & $0.67$ & $ 0.34$~\phantom{$-$}$(0.42)$ \\
\bottomrule
\end{tabular}
\end{table}


\FloatBarrier
 
\section{Conclusion}
\label{sec:conclusion}

 
This paper presents a search for dark matter produced in association
with a Higgs boson in final states with two hadronically decaying $\tau$-leptons and
large missing transverse momentum. The search is based on
a $139~\mathrm{fb^{-1}}$ dataset of $\sqrt{s}=13$~\TeV\ proton--proton
collisions recorded by the ATLAS experiment at the LHC during 2015--2018.
The analysis utilises two different signal regions which in turn
are divided into four and two $m_\text{T}^{\tau_{1}} + m_\text{T}^{\tau_{2}}$ bins, respectively, to target different
parts of the parameter space.
No significant deviations from the Standard Model prediction is found.
 
The results are interpreted in the context of the \TwoHDMa model where parameters are
varied to set $95\%$ CL limits in the $m_a$--$m_A$ and $m_A$--$\tan\beta$
planes.
Model-independent upper limits are set on the visible cross-section $\sigma_\text{vis}$ for eventual beyond the Standard Model physics processes producing a large missing transverse energy in association with a Higgs boson decaying to $\tau$-leptons.
These $95\%$ CL limits vary between 0.04 and 0.08~fb depending on which of the signal region bins is considered.
 
The presented work is the first exploration of the mono-Higgs signature with the Higgs boson decaying into a pair of hadronically decaying $\tau$-leptons with the ATLAS experiment. Compared to the $h+\met$ search with the Higgs boson decaying into a pair of $b$-quarks \cite{EXOT-2018-46}, this search offers sensitivity to the low $\tan\beta$ region of the $m_A$--$\tan\beta$ plane. In the $m_a$--$m_A$ plane this search has comparable sensitivity in the low $m_A$ region with an orthogonal selection, allowing for a combination of the two results.


\section*{Acknowledgements}


We thank CERN for the very successful operation of the LHC, as well as the
support staff from our institutions without whom ATLAS could not be
operated efficiently.
 
We acknowledge the support of
ANPCyT, Argentina;
YerPhI, Armenia;
ARC, Australia;
BMWFW and FWF, Austria;
ANAS, Azerbaijan;
CNPq and FAPESP, Brazil;
NSERC, NRC and CFI, Canada;
CERN;
ANID, Chile;
CAS, MOST and NSFC, China;
Minciencias, Colombia;
MEYS CR, Czech Republic;
DNRF and DNSRC, Denmark;
IN2P3-CNRS and CEA-DRF/IRFU, France;
SRNSFG, Georgia;
BMBF, HGF and MPG, Germany;
GSRI, Greece;
RGC and Hong Kong SAR, China;
ISF and Benoziyo Center, Israel;
INFN, Italy;
MEXT and JSPS, Japan;
CNRST, Morocco;
NWO, Netherlands;
RCN, Norway;
MEiN, Poland;
FCT, Portugal;
MNE/IFA, Romania;
MESTD, Serbia;
MSSR, Slovakia;
ARRS and MIZ\v{S}, Slovenia;
DSI/NRF, South Africa;
MICINN, Spain;
SRC and Wallenberg Foundation, Sweden;
SERI, SNSF and Cantons of Bern and Geneva, Switzerland;
MOST, Taiwan;
TENMAK, T\"urkiye;
STFC, United Kingdom;
DOE and NSF, United States of America.
In addition, individual groups and members have received support from
BCKDF, CANARIE, Compute Canada and CRC, Canada;
PRIMUS 21/SCI/017 and UNCE SCI/013, Czech Republic;
COST, ERC, ERDF, Horizon 2020 and Marie Sk{\l}odowska-Curie Actions, European Union;
Investissements d'Avenir Labex, Investissements d'Avenir Idex and ANR, France;
DFG and AvH Foundation, Germany;
Herakleitos, Thales and Aristeia programmes co-financed by EU-ESF and the Greek NSRF, Greece;
BSF-NSF and MINERVA, Israel;
Norwegian Financial Mechanism 2014-2021, Norway;
NCN and NAWA, Poland;
La Caixa Banking Foundation, CERCA Programme Generalitat de Catalunya and PROMETEO and GenT Programmes Generalitat Valenciana, Spain;
G\"{o}ran Gustafssons Stiftelse, Sweden;
The Royal Society and Leverhulme Trust, United Kingdom.
 
The crucial computing support from all WLCG partners is acknowledged gratefully, in particular from CERN, the ATLAS Tier-1 facilities at TRIUMF (Canada), NDGF (Denmark, Norway, Sweden), CC-IN2P3 (France), KIT/GridKA (Germany), INFN-CNAF (Italy), NL-T1 (Netherlands), PIC (Spain), ASGC (Taiwan), RAL (UK) and BNL (USA), the Tier-2 facilities worldwide and large non-WLCG resource providers. Major contributors of computing resources are listed in Ref.~\cite{ATL-SOFT-PUB-2021-003}.


\clearpage

\printbibliography
 
\clearpage
 
\begin{flushleft}
\hypersetup{urlcolor=black}
{\Large The ATLAS Collaboration}

\bigskip

\AtlasOrcid[0000-0002-6665-4934]{G.~Aad}$^\textrm{\scriptsize 102}$,
\AtlasOrcid[0000-0001-7616-1554]{E.~Aakvaag}$^\textrm{\scriptsize 16}$,
\AtlasOrcid[0000-0002-5888-2734]{B.~Abbott}$^\textrm{\scriptsize 120}$,
\AtlasOrcid[0000-0002-1002-1652]{K.~Abeling}$^\textrm{\scriptsize 55}$,
\AtlasOrcid[0000-0002-8496-9294]{S.H.~Abidi}$^\textrm{\scriptsize 29}$,
\AtlasOrcid[0000-0002-9987-2292]{A.~Aboulhorma}$^\textrm{\scriptsize 35e}$,
\AtlasOrcid[0000-0001-5329-6640]{H.~Abramowicz}$^\textrm{\scriptsize 151}$,
\AtlasOrcid[0000-0002-1599-2896]{H.~Abreu}$^\textrm{\scriptsize 150}$,
\AtlasOrcid[0000-0003-0403-3697]{Y.~Abulaiti}$^\textrm{\scriptsize 117}$,
\AtlasOrcid[0000-0003-0762-7204]{A.C.~Abusleme~Hoffman}$^\textrm{\scriptsize 137a}$
% ...
\bigskip
\\

% ...
$^{16}$Department for Physics and Technology, University of Bergen, Bergen; Norway.\\
% ...
$^{29}$Physics Department, Brookhaven National Laboratory, Upton NY; United States of America.\\
% ...
$^{35}$$^{(a)}$Facult\'e des Sciences Ain Chock, R\'eseau Universitaire de Physique des Hautes Energies - Universit\'e Hassan II, Casablanca;$^{(b)}$Facult\'{e} des Sciences, Universit\'{e} Ibn-Tofail, K\'{e}nitra;$^{(c)}$Facult\'e des Sciences Semlalia, Universit\'e Cadi Ayyad, LPHEA-Marrakech;$^{(d)}$LPMR, Facult\'e des Sciences, Universit\'e Mohamed Premier, Oujda;$^{(e)}$Facult\'e des sciences, Universit\'e Mohammed V, Rabat;$^{(f)}$Institute of Applied Physics, Mohammed VI Polytechnic University, Ben Guerir; Morocco.\\
% ...
$^{55}$II. Physikalisches Institut, Georg-August-Universit\"{a}t G\"ottingen, G\"ottingen; Germany.\\
% ...
$^{102}$CPPM, Aix-Marseille Universit\'e, CNRS/IN2P3, Marseille; France.\\
% ...
$^{117}$Department of Physics, New York University, New York NY; United States of America.\\
% ...
$^{120}$Homer L. Dodge Department of Physics and Astronomy, University of Oklahoma, Norman OK; United States of America.\\
% ...
$^{137}$$^{(a)}$Departamento de F\'isica, Pontificia Universidad Cat\'olica de Chile, Santiago;$^{(b)}$Millennium Institute for Subatomic physics at high energy frontier (SAPHIR), Santiago;$^{(c)}$Instituto de Investigaci\'on Multidisciplinario en Ciencia y Tecnolog\'ia, y Departamento de F\'isica, Universidad de La Serena;$^{(d)}$Universidad Andres Bello, Department of Physics, Santiago;$^{(e)}$Instituto de Alta Investigaci\'on, Universidad de Tarapac\'a, Arica;$^{(f)}$Departamento de F\'isica, Universidad T\'ecnica Federico Santa Mar\'ia, Valpara\'iso; Chile.\\
% ...
$^{150}$Department of Physics, Technion, Israel Institute of Technology, Haifa; Israel.\\
$^{151}$Raymond and Beverly Sackler School of Physics and Astronomy, Tel Aviv University, Tel Aviv; Israel.\\
% ...
$^{a}$ Also Affiliated with an institute covered by a cooperation agreement with CERN.\\
$^{b}$ Also at An-Najah National University, Nablus; Palestine.\\
$^{c}$ Also at Borough of Manhattan Community College, City University of New York, New York NY; United States of America.\\
$^{d}$ Also at Bruno Kessler Foundation, Trento; Italy.\\
$^{e}$ Also at Center for High Energy Physics, Peking University; China.\\
$^{f}$ Also at Center for Interdisciplinary Research and Innovation (CIRI-AUTH), Thessaloniki; Greece.\\
% ...

\end{flushleft}


\end{document}